# RECOMMENDATION

It is certified that Mr. **Ganesh Aryal** has carried out this dissertation work entitled

**"EFFECT OF SIZE DEPENDENT STRAIN AT DIFFERENT COVERAGE ON ISLAND FORMATION: A KINETIC MONTE CARLO STUDY"** under my supervision and guidance.

I recommend this dissertation in the partial fulfillment of the requirement for Master's Degree of Science in Physics.

……………………………….

**Prof. Dr. Shankar Prasad Shrestha**

Tribhuvan University

Department of Physics

Patan Multiple Campus

Patan Dhoka, Lalitpur

Nepal

Date: ........................................

# ACKNOWLEDGEMENT

I would like to express my sincere gratitude to my supervisor, **Prof. Dr. Shankar Prasad Shrestha** for his precious suggestions, extraordinary cooperation and encouragement throughout this work. It has been really pleasant and truly rewarding experience to work with him.

I would wish to express my hearty thanks to our professors and teachers, Prof. Dr. Bhadra Pokharel, Prof. Dr.Narayan Psd. Adhikari Prof. Dr. Jeevan Jyoti Nakarmi, Prof. Hom Nath Poudel, Dr. Leela Pradhan, Mr. Meen Prasad Dhakal, Mr. Basanta Raj Giri, for their unforgettable contribution and encouragement throughout the work.

I would like to extend the acknowledgement to my seniors Mr. Madan Raj Mul, my friends especially to Mr. Upendra Patel, Mr. Bhojraj Bhandari, Mr. Rajaram Ghimire, Mr. Aashish Poudel, Mr. Rameshwor Poudel Mr. Sagar Acharya and Bikram Acharya, for their constant supports and immense assistances to complete my work.

Finally, I must express my deepest gratitude and appreciation to my family and specially my wife Mrs. Laxmi Gautam for the inspiration and help me for the best carrier in my life.

**Ganesh Aryal**

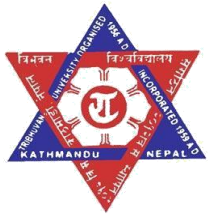

# EVALUATION

We certify that this thesis submitted to Dean Office, Institution of Science and Technology, Tribhuvan University, by Ganesh Aryal titled "**EFFECT OF SIZE DEPENDENT STRAIN AT DIFFERENT COVERAGE ON ISLAND FORMATION: A KINETIC MONTE CARLO STUDY**" meets all the pre-requisites in scope and quality as a dissertation in the partial fulfillment for the requirement of Master's Degree of Science in Physics, and has been approved by the undersigned members of the evaluation committee.

...................................................... ..................................................

**Prof. Dr. Shankar Prasad Shrestha**     **Prof. Hom Nath Poudel**
(Supervisor)     (Head of the Department, M.Sc. Physics)

Department of Physics     GoldenGate International College

Patan Multiple Campus     Battisputali, Kathmandu

Patan Dhoka, Lalitpur , Nepal     Nepal

………………………………     …………………………

**Internal Examiner**     **External Examiner**

Date:...............................

# ABSTRACT

Molecular beam epitaxy (MBE) is an epitaxy method for thin film deposition of single crystals. MBE is widely used in the manufacture of semiconductor devices, including transistors, and it is considered one of the fundamental tools for the development of the nanotechnologies. We have used Kinetic Monte Carlo (KMC) simulation technique to study the crystal growth during Molecular Beam Epitaxy (MBE). KMC is used to simulate the physical phenomena in which the probable events are selected by using pseudo random numbers. In this work, we have studied the effect of strain at various fixed coverage values, with the model which includes varying strain-driven detachment on average island morphology, average island density, relative width distribution and island size distributions. Our simulation results demonstrate that growth on a patterned impurity surface with inclusion of varying strain-driven detachment resulted in improvement the spatial ordering- on island morphologies and the size distribution of the islands.

# CONTENTS

# CHAPTER 1

## INTRODUCTION

Crystal growth is an interedispliniary field. It involves a variety of research fields ranging from surface physics, crystallography and material sciences to condensed matter physics. The self-organized growth of quantum dots lead to a growth of matter in patterned surface. It might help to simplify the fabrication of novel devices based on quantum dots, for instance, quantum dot lasers with better gain and lower threshold than conventional quantum well lasers [1-3]. So crystal growth acts as a bridge between science and technology for practical applications.

Epitaxial growth refers to depositing film on top of a crystalline substrate in orderly manner, replicating the atomic arrangement of the substrate. **Epitaxy** refers to the deposition of a crystalline over layer on a crystalline substrate. The over layer is called an **epitaxial** film or **epitaxial** layer. The term **epitaxy** comes from the Greek roots epi (ἐπί), meaning "above", and taxis (τάξις), meaning "an ordered manner". This epitaxy process is vastly used in nano-electronics as it is capable to grow high quality thin films with various compositions. The commonly used methods for epitaxial growth are Chemical Vapor Deposition (CVD), Vapor Phase Epitaxy(VPE), and Molecular Beam Epitaxy (MBE) [4]. Among the methods of the epitaxial growth, MBE takes a special place because it can grow films with desired specifications on the atomic level. Since it has a number of inherent advantages, such as: its flexibility, high cleanliness, possibility of conducting entire process of production in the vacuum, applicability of masks, ability of the independent variation of the growth rates and crystallization temperatures, and rapid switch from one deposited material to another[5]. So, it is an important technological process for fabrication of the nanostructures of high purity crystal. Through accurate control of the growth environment MBE enhances our ability to test theories of crystal growth with the help of modern surface characterization techniques, such as scanning tunneling microscopy[6,7]. A combination of experimental, theoretical, and simulation studies are now needed to fully understand the fundamental nature of film growth. For example, the precise control over the growth parameters and the actual determination of observables may still be difficult in experiments but are straightforward by the means of computer simulations.

In this epitaxial growth mode small islands which form quantum dots of the desired size arise spontaneously when a material B is grown on a lattice-mismatched substrate material A. During this growth mode one can not only observe the formation of the quantum dots, but one also finds some self-ordering in the position and the size of those. Under special growth conditions equally sized quantum dots may arrange themselves in a regular square lattice. The strain induced by the quantum dots in the substrate is of great importance for the ordering[8]. The self-organization is manifested on different levels. There is the regular shape of the quantum dots, their size distribution, and the formation of regular arrays of quantum dots on the substrate.

Monte Carlo method refers to a broad class of algorithms that solve problems through the use of random numbers and probability statistics to investigate the problem. The most famous of the Monte Carlo methods is the Metropolis algorithm, invented just over 50 years ago at Los Alamos

National Laboratory[9]. This is accomplished through surprisingly simple rules, involving almost nothing more than moving one atom at a time by a small random displacement. In general, to call something a "Monte Carlo" method, all one need to do is use random numbers to examine ones problem. In 1960's researchers began to develop a different kind of Monte Carlo algorithm for evolving systems dynamically from state to state. The ordinary Monte Carlo cannot be used in the study of time dependent quantities. Kinetic Monte Carlo(KMC), on the other hand, is a method for solving the kinetic equations[10]. The popularity and range of applications of KMC has continued to grow and it is now a common tool for studying materials subject to irradiation and surface growth. The purpose of the KMC is to reflect the time evolution of the system and to reproduce the non-equilibrium process. Therefore, KMC is well suited for the modeling of the different growth phenomena. The idea behind the KMC is to describe the diffusive motion of atoms on a surface under the periodic atomic potential of the surface and other atoms adsorbed on the surface. Each atom can move from its actual position to another lattice site, hereby influencing other atoms on the surface. Here we restrict ourselves to only nearest-neighbor hopping and nearest-neighbor interactions. Hence each atom has a certain probability p per unit time to move from its position to a neighboring site; this probability is proportional to an Arrhenius fact[11],

$$P = \nu \exp(-(E_s + nE_n/K_BT)) \qquad (1.1)$$

Where $E_s$ is the atomic binding energy to the surface and $E_n$ is the binding energy to a single nearest-neighbor atom, n gives the number of nearest neighbors, T is the temperature and $K_B$ Boltzmann's constant. $\nu$ is vibration frequency.

Strain is induced due to lattice mismatch by the deposited ad-atoms on the substrate, for instance the lattice parameter of Silicon is 0.54310nm and the lattice parameter of germanium is 0.56575nm. The lattice parameters of germanium and silicon have a 4.17% mismatch. The lattice mismatch ($f_m$) is calculated from equation (1.2):

$$f_m = (a_s - a_l)/a_l \qquad (1.2)$$

where, $a_s$ is the lattice parameter of the substrate and $a_l$ is the lattice parameter of the layer[18]. In strained systems the equilibrium position and also the equilibrium binding energy of the atoms will be distorted so it is difficult to properly include strain in simulation of epitaxial growth for systems of reasonable size. The reason is that solving the elastic equations is rather expansive. It is almost prohibitively expensive to solve the elastic equations at every time step in an atomistic simulation, such as kinetic Monte Carlo simulation[12]. Thus, the binding energy in strained systems will in general be lower than in the unstrained case. To include this effect we introduce a correction term $E_c$ for the binding energies $E_s$ and $E_n$. $E_c$ may be different at each atomic position (x, y) on the substrate surface, because of the non-homogeneous strain field. The resulting hopping probability is [13]:

$$P = \nu \exp(-(E_s + nE_n + E_c)/K_BT) \qquad (1.3)$$

In this context Ec is always negative to reduce the binding energy and to increase the hopping probability. The higher the strain the more negative is Ec. As a result atoms will move faster in areas with higher strain and slower in areas with lower strain and this will lead to a flux of atoms from high-strain to low-strain regions. To obtain the strain field in general one has to use the theory of elasticity, which was utilized, e.g., in Ref.[14] for an energy balance, but here for simplicity we will employ some phenomenological functions to mimic the strain fields Ec of island. In the first step the hopping probabilities for all atoms on the surface are calculated. For the binding energies we use Es=0.4ev and En=0.1ev. Atoms with four nearest neighbors are considered as immobile. To incorporate the correction Ec due to the strain field we take all island with minimum size (in the simulations we use, e.g., island with at least N=10 atoms), and set the range of strain field around those islands equal to twice their radius. We use the relation Ec= $\alpha$ *N to find the strain energy of the each islands. Where $\alpha$=0.01ev[15]. In this dissertation, we focus the effect of patterned impurity on the surface with strain-driven detachment using Kinetic Monte Carlo simulation, there exists an optimal value of flux at which most uniform islands are formed. The pre-deposition of impurities as pattern and the strain enhanced the detachment rates of edge atoms from other than impurity nucleating islands for growth of impurity nucleating islands and strain monomer diffusion responsible for size ordering of the growth impurity nucleating islands on substrate is the key part of our work. We have studied its effect on the morphology of island, average island size, average island density, relative width distribution and island size distribution with fixed values of energy parameters and various values of flux.

The outline of this dissertation is summarized as follows: In chapter-2, the basic concepts related to crystal growth are discussed. These include issues such as atomic processed during growth, formulation of growth models, and surface diffusion and nucleation theory. Chapter-3, describes a general introduction to the Monte Carlo method with a more detailed description of the Kinetic Monte Carlo method and its implementation. Chapter-4, presents the results obtained in this work and the results are also discussed and analyzed. Finally, Chapter-5, is summary of the performed work and suggestions for future works.

# CHAPTER 2

## 2.1 Surface growth

To understand and control over the surface growth of epitaxial thin films, experimental methods such as x-ray diffraction, scanning tunneling microscopy (SEM), and transmission Electron Microscopy (TEM) as well as several theoretical and computer simulation tools such as Molecular Dynamics (MD), Monte Carlo (MC) and Kinetic Monte Carlo (KMC) simulation have been widely used. Particularly, KMC is a well suited method to study surface growth. The method could predict and model a crystal growth process in large time scale from micro-second to hours.

Solid interface is defined as a small number of atomic layers that separate two solids in close contact with one another[16]. For instance, the surface of a solid is a particularly simple type of interface, at which the solid is in close contact with the surrounding atmosphere or, in the ideal case, the vacuum, so the properties of the interface differ significantly from those of the bulk material it separates. The macroscopic properties of a surface are closely associated with the microscopic details determined by the atomic structure. The term morphology refers to the form of shape of a surface in a more macroscopic sense. The scale, however, largely depends on the type of property being considered. Surface structure, on the other hand is associated with a microscopic, atomistic picture. It is used to denote the detailed geometrical arrangement of atoms and their relative positions in space. It should be noted that these two concepts are closely related as the atomistic structure significantly influences the morphology of a surface. Solid interfaces between thin films and solid substrates are of major importance in modern technology. The focus of this study is on epitaxial growth.

In epitaxy, a solid interface advances through the addition of new material. The deposit crystallizes in a manner dictated by the lattice structure of the underlying solid and the crystal structure of the substrate is retained up to a certain thickness of the growing deposit. The term homo-epitaxy is used when two crystals are identical and if are not refer to hetero-epitaxy. In order to understand the process of thin film growth, we considered the underlying principles that determine the structure and morphology of a particular film in an atomistic point of view. Thus, in this work, we aim to build a model of epitaxial growth based on the description of the individual atomic processes involved in film growth with considering variable strain that produce due to lattice misfit.

### 2.1.1 Growth as a Non-equilibrium Process

As already mentioned above, the macroscopic properties of a surface are closely associated with the microscopic details determined by the atomic structure. In thermodynamic equilibrium, all atomic processes proceed opposite directions at equal rates, as infer by "detailed balance". This is possible only when adsorption and formation of clusters at the same rate as their decay. Hence, in equilibrium there is no net growth and average macroscopic quantities such as surface coverage and roughness stay constant. The various microscopic surface processes then cause fluctuations around these equilibrium quantities which are usually well described by Statistical Mechanics. Therefore, in order to obtain net growth rate one has to be far from equilibrium condition. The degree to which growth proceeds away from equilibrium decides to which extent the morphology will be determined by thermodynamic quantities, such as surface and interface free energies, or by growth kinetics. The driving force for crystal growth is the difference of chemical potentials of the crystal and vapor phase, $\Delta\mu= \mu v – \mu c$. It is called the *supersaturation* [17].

The growth on far from equilibrium typically characterizes the situation encountered in a technique, called Molecular Beam Epitaxy (MBE), where a high supersaturation ($\Delta\mu$) is applied in order to grow the films rapidly from the vapor phase. Under such conditions, growth is to a high degree determined by kinetics and the microscope path taken by the system becomes decisive. Understanding and controlling the growth morphologies in the kinetic regime therefore requires detailed knowledge of the microscopic processes involved like atomic diffusion events. These atomic displacements are thermally activated and their rates are generally well described by Boltzmann statistics. Measurements of the activation barriers and pre- factors are required to determine the hierarchy of rates and thus to identify the rate limiting steps. Knowledge on these energetic barriers in turn allows one to predict and influence the island and film morphologies.

### 2.1.2 Epitaxial Growth

Epitaxial growth refers to depositing film on top of a crystalline substrate in orderly manner, replicating the atomic arrangement of the substrate. **Epitaxy** refers to the deposition of a crystalline over layer on a crystalline substrate. The over layer is called an **epitaxial** film or **epitaxial** layer. The term **epitaxy** comes from the Greek roots epi (ἐπί), meaning "above", and taxis (τάξις), meaning "an ordered manner". It is one of the most important and versatile methods of crystal growth for device application. To produce semiconductor structures for microelectronic devices, the initiating point is a single-crystal wafer such as silicon. The desired materials, atoms or molecules, are than deposited on top, with the added atoms conforming to the underlying crystal structure. This process during which a crystal layer is formed on an underlying crystalline surface, substrate, as a result of deposition is called epitaxial growth.

### 2.1.3 Kinetics- Atomic process during growth

To understand the crystal growth phenomena, the detailed knowledge of the atomic processes is required. The basic atomic processes taking place during growth on the crystal interface are the following: *deposition* of the ad-atom on the surface, *surface diffusion*, and *desorption* from the surface. Growth is initiated by introducing a flux of incoming particle on to an initially flat surface. Typically, the deposition material is thermally evaporated from a source, forming a beam of neutral atoms or molecules that is directed to the growing surface. A process contending with deposition is desorption. The desorption probability depends on two factors, the energy barrier ED, that the atom needs to overcome in order to leave the surface, and the temperature T. The average time an atom remains on the surface follows an Arrhenius-type of law [18],

$$\tau= \tau_o \exp (-E_D/K_B T) \qquad 2.1$$

where, the pre-factor $\tau_o$ is of the order of $10^{-14}$ s and $K_B$ is the Boltzmann constant. The desorption rate increases if the temperature is increased eventually reaching a point where the surface begins to evaporate.

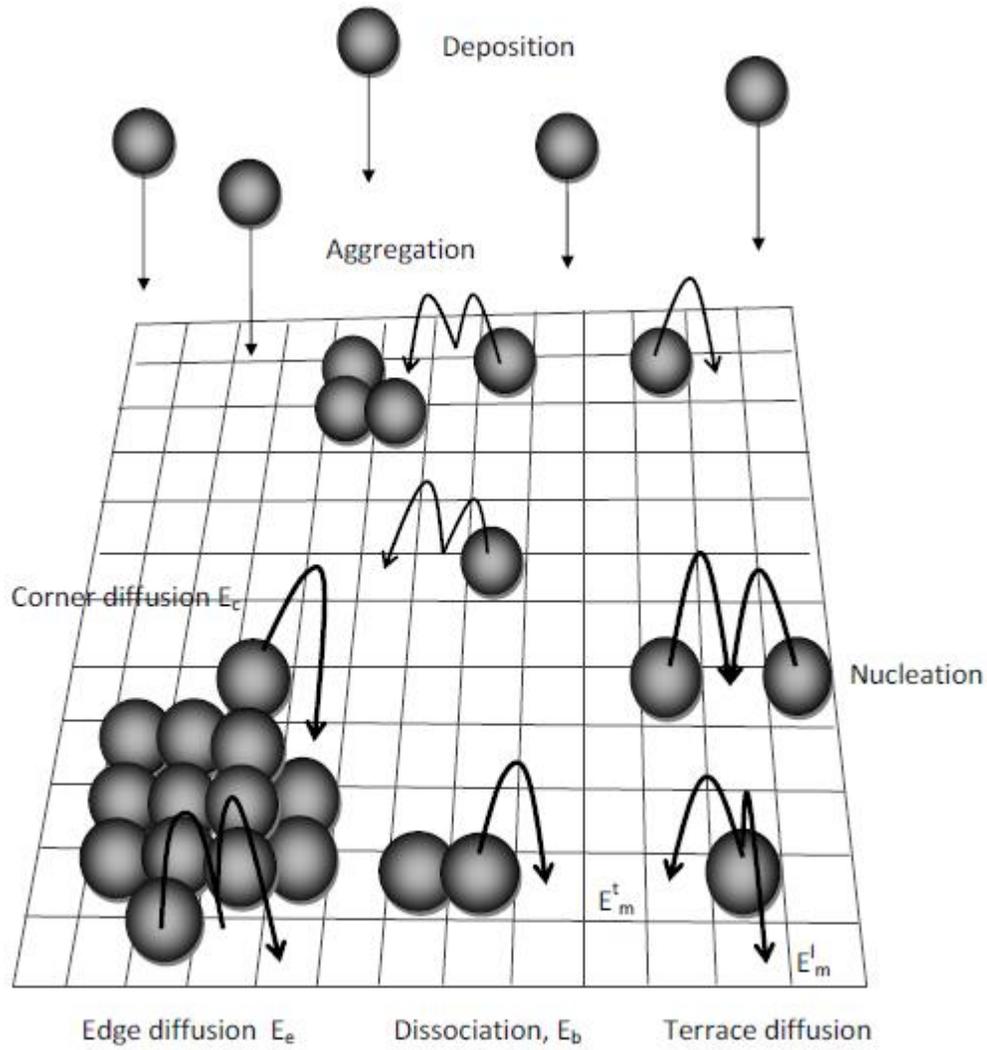

Fig.2.1 Atomic process in the kinetics of 2-D films growth.

## 2.2 Diffusion on the Surface

During epitaxy, atoms arrive via the molecular beam and adsorb to the surface. The atoms that have been adsorbed to the surface are called adatoms. These adatoms diffuse across the surface and incorporate into the crystal, forming another layer of the periodic structure. One layer of atoms equal to $6.8 \times 10^{14}$ per $cm^2$ for Si, is called a monolayer (ML). The amount of material added to an initial substrate is termed as coverage and is typically measured in units of ML. Small value of coverage generally less or equal to 10% corresponds 2D islands otherwise 3D islands [19]. Once a layer is completed, there may continue to be some inter-diffusion but this occurs at a much slower rate than surface diffusion [20].

The migration of single adatom on the surface is the most important diffusion process in the growth of thin films, which leads to raise the formation of islands on the surface. These adatoms will diffuse on the surface with some probability. Individual adatoms are located on the surface at a minimum energy adsorption site. They may undergo transitions between two adjacent adsorption sites if the energy provided by thermal fluctuations of the lattice is large enough i.e. the temperature is sufficiently high. Thus, depending on the substrate temperature, the adatoms migrate along the surface by overcoming the activation barrier $E_m$ at the saddle point between neighboring sites. The hopping image for the transition is only valid for $K_B T \ll E_m$, which means that the residence time spent at the adsorption site is long in comparison with the time spent in the transition state. Between jumps, the adatoms therefore looses all memory of its original direction, and thus a random migration evolves. For $K_B T \gg E_m$ the hopping image is no longer valid as no energetic barrier hinders the lateral displacements of adatoms. The adatoms move on the surface by diffusion events, such as terrace, interlayer, corner and edge-diffusion[3,27-28]. These atomic events are thermally activated and their rates depend on the local surroundings and can usually be described by Boltzmann statistics.

The diffusion rate is defined as the average number of jumps an adatom takes in a unit time interval. The hopping rate of a free surface adatom has exponential temperature dependence, given by the Arrhenius expression

$$D_o = k_o \exp(-E_m/K_BT) \quad (2.2)$$

Where the pre-factor $k_0 = 2K_BT/h$ is the frequency of small atomic vibrations, h is Planck's constant.

## 2.3 Growth Classification

The growth is usually classified in three modes based on thermodynamic arguments [16]:

(a) Frank-Van der Merwe or Two Dimensional Growth Mode

(b) Volmer-Weber or Three Dimensional Growth Mode

(c) Stranski-Krastanov or Intermediate Growth Mode

The simplest growth mode, Frank-Van der Merwe growth mode is two dimensional layer by layer growth, where one complete monolayer grows after the other. Such a 2D growth is preferred if the sum of the surface free energy ($\gamma A$) of the adsorbate and its interface free energy ($\gamma AB$) is larger than the substrate free energy ($\gamma B$).
i.e. $\gamma A + \gamma AB > \gamma B$ \quad (2.3)
Conversely, if the sum of surface free energy of the adsorbate and its interface free energy is less than the substrate free energy; the growth doesn't form monolayers. It rather forms little 3D droplets very similar to water droplets on a freshly sealed car top. This growth mode is referred to as the Volmer–Weber growth mode.
i.e. $\gamma A + \gamma AB < \gamma B$ \quad (2.4)

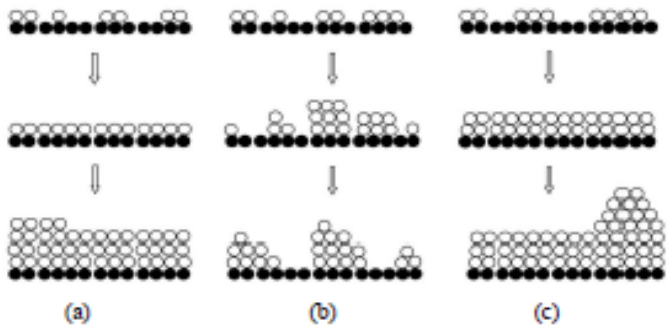

Fig. 2.2: Modes of growth; (a) Frank-van der Merwe, (b) Volmer-Weber, (c) Stranski Krastanov.

There is an interesting case, named Stranski-Krastanov growth where the solid droplets are not quite like liquid. They come up with additional features like well defined lattice constant and certain elastic properties. Thus, in the phenomenon of heteroepitaxy, a lattice misfit between the substrate and the adsorbate causes the growing layers to be strained. In this situation, it is energetically favorable for the first few layers to grow in layer by layer fashion, but after that the strain is relieved more efficiently in 3D mode. This is an intermediate case, between the 2D layer by layer growth mode and Volmer–Weber growth mode.

The growth mode changes with substrate temperature, or with the strength of the disequilibrium. In far from equilibrium MBE growth, the growth mode is usually changes from 3D growth at low temperatures to layer by layer growth at higher temperatures. It is because at higher temperatures, the adatoms on the surface becomes so mobile that they can jump over local energy barriers, including step barriers, which at low temperature, prevent an atom from jumping down a step (a necessary process for layer growth) and the surface grows in 3D mode[21] . Larger lattice mismatches generally favor island-like growth, and higher deposition rates, on the other hand, lead to a greater tendency towards layer-like growth. Under certain conditions, when the adatoms are mobile but the barriers at step edges suppress the inter-layer transport, the growth results in the creation of large, ordered, pyramid-like formations on the surface. This is called the unstable MBE growth mode. It was first detected by Villain in 1991[29], and later studied by computer simulations[30,31]. Recently experimental findings of various systems have also been reported: for example GaAs/GaAs(001)[32].

## 2.4 Models of crystal growth

There are two complimentary approaches in the study of crystal growth[22]. In atomistic or discrete approaches the position of every atom is well defined, while continuum approaches view the interface on a coarse-grained scale, in which every property is averaged over a small volume containing several atoms. The resolution of experimental techniques has developed rapidly over the past years. Scanning tunneling microscopy (STM) and atomic force microscopy (AFM) are techniques capable of indentifying not only the structure of the surface but the positions of individual atoms as well. Based on this detailed information, modeling of growth processes on the atomic level is becoming a widely used tool to gain understanding of the underlying mechanisms. The predictive power of the continuum approaches is limited to length scales larger than the typical inter-atom distance. The purpose of modeling is to concentrate on a few presumably important aspect of the process in question and disregard other details. Because of this, the model is always an approximation. However, computer simulations offer a possibility to study the influence of several mechanisms under well-defined conditions. A growth model consists of two parts: a geometrical and dynamical part[23]. In the widely used lattice gas model, the system of solid and vapor are described by a lattice with some sites occupied and some left empty. The particles can occupy only discrete position in space. The regions of high concentration of atoms correspond to the crystal phase and the regions of low concentration to the gas phase [Fig. 2.3(a)].

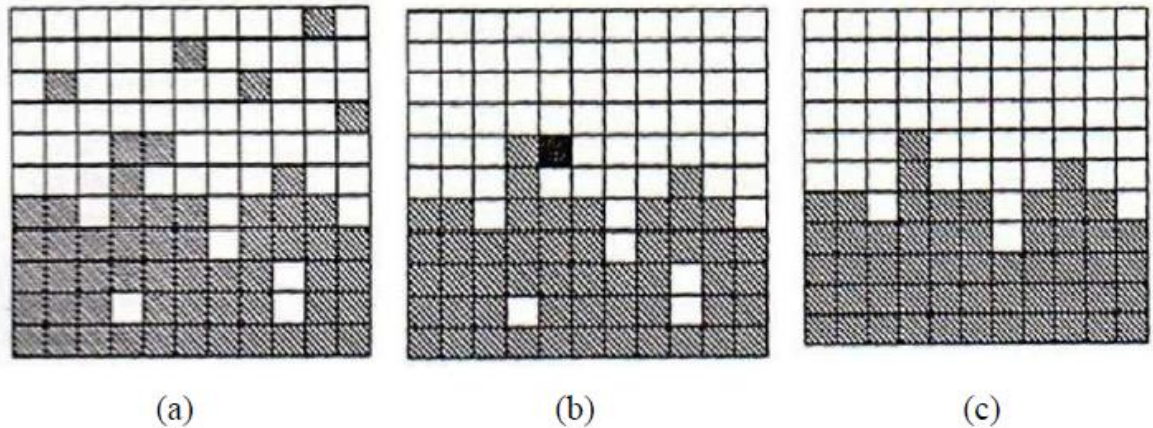

Fig. 2.3: Hierarchy of approximation in discrete growth models illustrated in the case of two dimensions. (a) The configuration of lattice gas with the bottom region corresponding to solid and the top region corresponding to gas. Atoms are represented by hatched squares. (b) Only positions of atoms of the solid are considered, new atoms are added by growth rules defined in a model. The darker square indicated the overhang. (c) Surface configurations in the SOS approximation.

However, the lattice gas model can be used to describe the crystal to a good approximation, and it is still acceptable (although not very good) for the gas phase. In the simulations performed in this study, some additional approximations to the model are made. A simple cubic lattice is used instead of the real structure of the material. This makes the technical realization of the simulation a lot simpler, and although some effects may be lost, the essential features should still be preserved by this approach. The bulk diffusion inside the crystal can also be neglected, because the interest lies in the motion of the atoms forming the growing surface. This applies well in the case of MBE growth where surface is directly bombarded by atoms. Moreover, the processes inside the fluid are also neglected. Thus the growth can be well described only by processes at the interface between the fluid and solid [Fig. 2.3 (b)].

In this picture, we are interested only in movements of atom representing the solid and the growth proceeds by adding atom at position determined by growth rule. This is well justified approximation for ballistic growth where mutual interactions of particles in the fluid are negligible. In addition to these approximations the so-called solid-on-solid (SOS) model[24] can be used when growth occurs on a planar surface and under conditions where, very few vacancies or overhangs are not allowed in the growing material [Fig. 2.3 (c)]. It means, each atom is sitting on the top of the other atom. In this case, the surface is described by a single valued height function, $h(x)$, of the substrate coordinate. This model is applied in all the kinetic Monte Carlo simulations performed in this study. The dynamical growth rules incorporate the second part of the growth model where some approximations are needed as well.

The motion of the individual particles is assumed to be instantaneous, independent, and Markovian. In order to develop a minimal model one select only process which are supposed to be relevant for the phenomena to be studied, and other deemed inventions are ignored. Each process takes place at a certain rate and these events are realized in the model at frequencies proportional to respective rates. The basic elementary processes, after the omission of bulk diffusion, are as follows:

(a) Deposition of an atom on the surface
(b) Diffusion of an adatom on the surface
(c) Attachment of an atom on the island and
(d) Detachment and Desorption of an atom from the surface

The rules of elementary processes are very little known and they are guessed with the use of all available experimental and theoretical information. They often depend on the local surroundings.

# CHAPTER 3

## SIMULATION METHOD

Computer simulation have been widely used in the field of crystal growth, at the same speed as the computational power is exploding, the number of simulation methods and applications spreads [25]. A challenge to perform „realistic" growth simulations is the wide range of relevant length and time scales. KMC simulations have been widely used in modeling epitaxial growth. This is because they allow easy implementation of a wide range of atomistic kinetic process whose rates can in theory be determined from first principles [26]. In classical condensed matter physics, two kinds of simulations have been proposed useful: Molecular Dynamics (MD) simulation and Monte Carlo (MC) simulation.

In this chapter, we shall briefly introduce MD simulation and the general algorithm associated to MC simulations and use KMC to construct atomistic two-dimensional models for the growth of islands.

## 3.1 Molecular Dynamics Simulation

Molecular Dynamics (MD) is a computer simulation method for studying the physical movements of atoms and molecules, and is thus a type of N-body simulation. The atoms and molecules are allowed to interact for a fixed period of time, giving a view of the dynamical evolution of the system. In the most common version, the trajectories of atoms and molecules are determined by numerically solving Newton's equations of motion for a system of interacting particles, where forces between the particles and their potential energies are calculated using interatomic potentials or molecular mechanics force fields. These simulations follow the evolution of a system by tracking atom movement in response to inter-atomic forces. From the setting of initial conditions, Newton's equations of motion are integrated forward with time to yield smooth movement. This results in exact positions and velocities of all the particles in the system. The interaction force describes the nature of used potential. The parameters of the potential are extracted from the parameters of the potential are extracted from the first principle calculations of quantum mechanical arguments or from experimental measurements. Experimentally measurable quantities are compared with simulation results and the parameters are adjusted until sufficient agreement between simulation and reality is achieved. The molecular dynamics simulation method is well suited for the study of time–dependent phenomena. The basic idea to perform Molecular Dynamics simulation is that at first for a system of interest, one has to specify a set of initial conditions (e.g. initial positions and velocities of all the particles in the system) and interaction potential for driving the forces among all the atoms. Then, a set of classical equation of motion for all particles are solved in the system and the evolution of the system in time is determined.

For simplicity, we begin by considering the NVE ensemble in which the number of particles N, the system volume V and the total energy E remain constant. The coupled equations of motion may be derived for example from Lagrange's or Hamilton's equations. In the former approach, Lagrange's equations for the system of N particles produce a set of 3N equations to be solved [27]:

$$m_i \ddot{r}_i = F_i = \nabla_{r_i} U(r^N), \qquad (3.1)$$

Where, $m_i$ is the mass of $i^{th}$ particle, $F_i$ is the total force acting on it, and $U$ is the appropriate inter-atomic potential. In the latter approach, where Hamilton's equations are used to derive to the dynamics of the system, we obtain the following set of 6N equations:

$$\dot{r}_i = P_i/m_i$$
$$\dot{P}_I = F_i \qquad (3.2)$$

Where, $p_i$ is the momentum of the particle. In this approach, the energy of the system is invariant with time so that solving these equations would produce a sequence of states in the micro-canonical ensemble. Either set of equations can be solved by finite difference methods using a time interval $\Delta t$ which must be made sufficiently small for accurate results.

The solutions of the equation (3.2) give the positions $r_i(t)$ and velocities $v_i(t)$ of all atoms as a function of time. Once the initial conditions and the interaction potential are defined, the equation of motion can be solved numerically.

The basic Molecular Dynamics algorithm consists of the following steps:

I. Initialization of the system

II. Computing the forces

III. Integrating the equations of motion

IV. Measurement of average quantities

An appealing feature of the MD method is that it follows the actual dynamical evolution of the system and no predictions need to be made about the possible trajectories. Comparing to KMC, to be discussed later, the MD method does not suffer from any of the problems related to forming a catalog of possible events or determining the correct rates for those events. However, the need to integrate the equations of motion dictates that the maximum time step is in the femto-second range at best, so that the problems which involve a much larger time scale cannot be addressed [32]. This is the drawback of MD that it is extremely computational intensive. To avoid errors in numerical integration, very small time steps must be taken. This step size is dictated by the fastest changing component of the simulation, which is especially problematic if one component (e.g. particle vibration) is much faster than the component of interest (e.g. particle diffusion). For example, a MD simulation involving a few hundred particles could take some days to simulate a few nanoseconds of movement, or millions of years to simulate a single second[27]. Recently, several new "accelerated dynamics methods" have appeared which aim to circumvent the time scale problem. The common feature of this class of methods is that they all try to speed up the natural dynamics of the system while retaining a sequence of states which is representative of the original dynamics. Further discussion is beyond the scope of thesis, but an excellent review of this subject is given by Voter *et al* [33].

## 3.2 Monte Carlo Simulations

Monte Carlo methods (or Monte Carlo experiments) are a broad class of computational algorithms that rely on repeated random sampling to obtain numerical results. Their essential idea is using randomness to solve problems that might be deterministic in principle. They are often used in physical and mathematical problems and are most useful when it is difficult or impossible to use other approaches. Monte Carlo (MC) refers to a group of methods in which physical or mathematical problems are simulated using random numbers [26]. These random numbers model the effect of fluctuations in the system. They have been commonly used in the computation of average quantities at thermodynamic equilibrium. For such problems, one is usually required to compute the partition function Z, which for systems with a large number of interacting particles can be a very difficult operation. In 1953, Metropolis et al. [34,35] developed the first MC method that solved this problem without direct computation of the partition function. The essence of MC lies in sampling. Instead of examining all, we pile up samples and obtain aimed quantity from the samples through an averaging procedure. The sampling is carried out in a stochastic manner with the aid of random numbers. So, the things of stochastic character are best suited for the application of the MC method. Two slightly different types of Monte Carlo simulations are common, in general, in computer simulation:

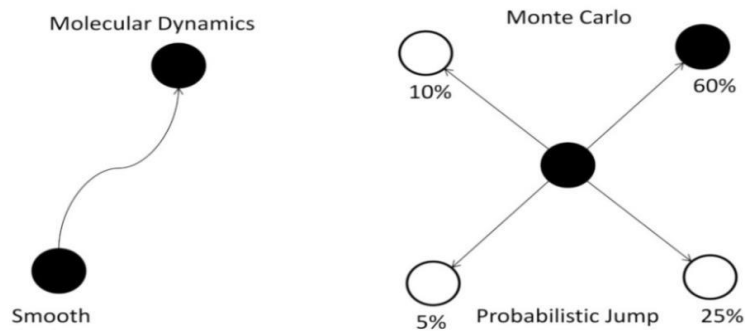

Fig. 3.1: Difference between MD and MC simulations

### 3.2.1. Standard Thermo-dynamical Monte Carlo

This type of simulation is useful for finding the equilibrium state of a system. It is not necessary to know the details of the time evolution of the system, or the precise traces of the all atoms in the system. The only thing of interest of the mean values of some observables: how often on average the deposited atom jumps to the neighboring sites, how long on average the deposited atom migrates before re-evaporation, etc.

At each step of the simulation, a particle is chosen and moved randomly to a trial position. Then, the total system energy of the resulting new state is calculated. If the total energy decreases, the trial movement is accepted and the state is updated. If the total energy increases, then the trial movement may still be taken but only with a certain probability which decreases the increased energy [23, 36]. Otherwise the trial movement is rejected. This algorithm is repeated until more favorable states are no longer encountered. Importantly, this type of simulation doesn't depend on time and is therefore unsuitable for kinetic process such as epitaxial. Still, it is very useful for performing relaxation in simulations.

The Monte Carlo method produces an essence by sampling and an aimed quantity is given as the ensemble average. Let us consider a system at a given temperature, or more precisely speaking a system which is in contact with a heat bath kept at a given temperature. Here, the elements of the ensemble are microscopic states $x_i$ of the system and the quantity A is given as the mean over the set of states.

$$<A> = \frac{1}{n}\sum_{i}^{n} A(x_i) \qquad (3.3)$$

How to sample the elements is a key point of the MC method. If we sample the elements randomly in the case of a constant temperature, equation (3.3) has no sense, because some states in equation (3.3) occur quite often and other states in equation (3.3) occur quite rarely. The essence of states must have a proper distribution in accordance with the condition of constant temperature. The ensemble of the states is generated by the Markov chain. The Markov chain is a generalized idea of time evolution of the state of Newtonian mechanics [37]. The Markov chain of states is constructed by generating from the just previously generated state:

$$P : x_0 \ldots\ldots\ldots\ldots x_n \qquad (3.4)$$

The state $x_i$ is derived from the state $x_{i-1}$ and there is no causality between them. But the state $x_i$ follows with the state $x_{i-1}$ with a probability $W(x_i, x_{i-1})$.

There are some possible choices for the formula of W. The most famous is the Metropolis function given by [27];

$$W(x', x) = \frac{1}{\tau}\exp\left(-\Delta H / K_B T\right), \qquad \text{if } \Delta H > 0$$

$$= 1/\tau, \text{ otherwise} \qquad (3.5)$$

Where $H$ is the Hamiltonian of the system, and $\Delta H = H(x') - H(x)$.

The Hamiltonian used in the MC contains only the configurational part of the usual Hamiltonian. Even if the kinetic part is induced, this part is cancelled when the Boltzmann's factor is normalized to express the absolute value of probability.

By using the Metropolis function, we can generate a Markov chain of states which has a correct distribution of a canonical ensemble [23, 36]. The actual procedure of the generation of states is given by the following algorithm.

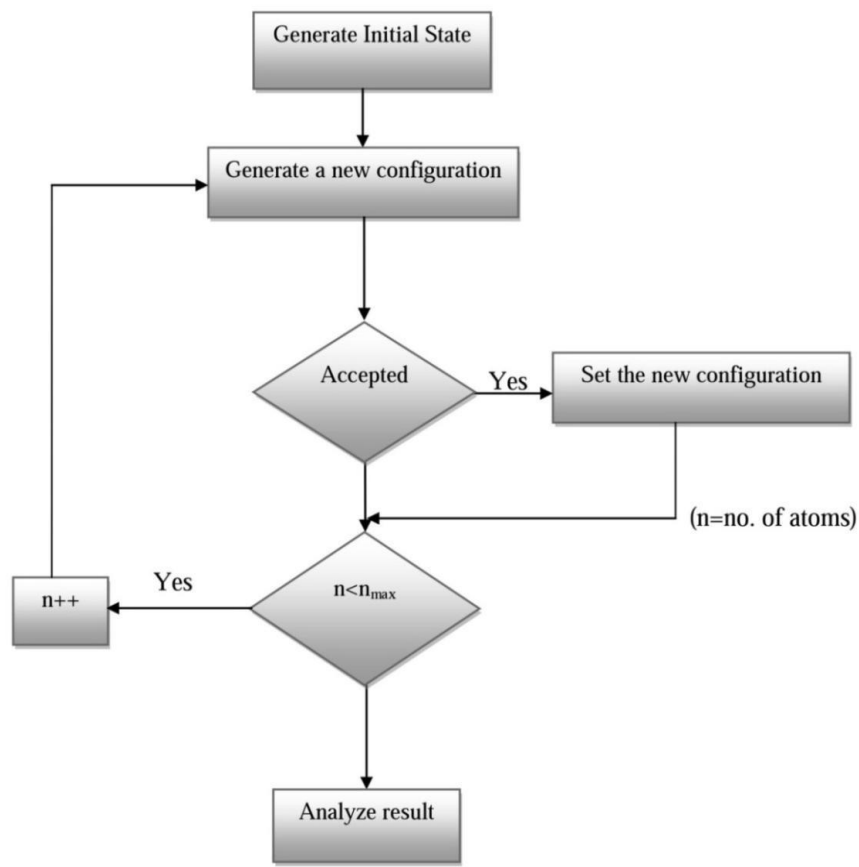

Fig. 3.2: A general Monte Carlo algorithm

**Algorithm 1**

1) Draw a uniform random number r which is between 0 and 1.

2) Using this number, choose a new state x'.

3) Compute the change in energy: $\Delta H = H(x') - H(x)$

4)        If $\Delta H \leq 0$, accept the new state x' and go to 1.

5)        If $\Delta H > 0$, draw a random number r.

6)        If $r < \exp\{-\Delta H / K_B T\}$, accept the new state x' and go to 1.

7)        Otherwise, reject the new state x' and the old state x is used again.

8)        Go to 1.

This is repeated until some convergence criteria are fulfilled and at the end, the results are analyzed.

It is important to note that the applicability of MC to dynamic phenomena can be examined by using master equation method. Thus, the MC method can be viewed as solving the master equation;

$$\frac{\partial P(x,t)}{\partial t} = -\sum_{x'} W(x',x)P(x,t) + \sum_{x'} W(x,x')P(x',t) \tag{3.6}$$

Where, P(x,t) is the probability distribution of the configurations at the time t.

$$\frac{\partial P(C)}{\partial t} = 0 \tag{3.7}$$

$$\Rightarrow W(x',x)P(x) = W(x,x')P(x') \tag{3.8}$$

The unique equilibrium distribution of the Markov chain satisfies the stationary condition. This is called the condition of detailed balance and necessary condition for the convergence of Markov chain. Fig. 3.2 above illustrates schematically how a MC simulation proceeds. A straightforward application of above algorithm couldn't be expected, in general, to describe correctly the dynamic properties. In the calculation of static quantities, the configurational changes do not need to correspond to physical events.

## 3.2.2 Kinetic Monte Carlo Simulation

A second type of probabilistic simulation is Kinetic Monte Carlo simulation. The **kinetic Monte Carlo (KMC)** method is a Monte Carlo method computer simulation intended to simulate the time evolution of some processes occurring in nature. Typically these are processes that occur with known transition rates among states. It is important to understand that these rates are inputs to the KMC algorithm, the method itself cannot predict them. In this case, the configurational changes have to correspond to the real events. Each of the events can happen with some probability per unit time (rate). That is this algorithm is explicitly time dependent through the use of event frequencies. Here, the steps are not taken completely at random but rather are based on the probability of certain events. For example, the probability of a chemical reaction occurring in a system depends on temperature, activation energy, and the frequency of reaction opportunities according to the Arrhenius equation,

$$p = \nu \exp\left(-E_{act}/K_B T\right) \qquad (3.9)$$

Evolution of a system is simulated by, at each time step, randomly choosing one of the possible events based on the relative probabilities of all events. The probabilities are expressed in per second units, derived from the frequency of reaction opportunities. After each step, the simulation clock advances by the inverse of the sum of all event probabilities.

Thus, KMC is the method for solving the master equation (3.6) above for the dynamic phenomena. But in case of KMC method, the condition of detailed balance doesn't need to be obeyed necessarily but it has been observed that it is advantageous in the case of diffusion. Also, when studying relaxation towards equilibrium, the detailed balance condition should be satisfied in order to guarantee convergence to the equilibrium distribution.

To be more specific for our problem of crystal growth, consider N be the total number of possible events in a given configuration C. These events are the possible atomic processes that are included in the model, the deposition or the diffusion of an ad-atom. Each event takes place in a certain rate $R_a$; $a = 1, 2, \ldots N$. Both the N and the rates $\{R_a\}$ depend on the configuration C. The total rate is defined by,

$$Q = Q(C) = \sum_{a=1}^{N} R_a \qquad (3.10)$$

The transition rate can now be written as,

$$W(C \to C') = \sum_{a=1}^{N} R_a V^a(C \to C') \qquad (3.11)$$

Where the stochastic matrix $V^a(C \to C')$ specifies whether the transition $(C \to C')$ can be realized by event *a*. Every process *a* is thermally activated with relative probability,

$$P_a = \frac{R_a}{Q(C)}.$$

The possible events are selected with probabilities proportional to their physical rates in the simulation.

The KMC method can be implemented in a straightforward way but the simplest algorithms often suffer from very low acceptance rates, which, in practice, make them unsuitable for many simulations. Therefore, two versions of a much faster algorithm without unsuccessful attempts are described here. One of the widely used algorithms in the simulation of crystal growth is BKL algorithm formulated by Bortz, Kalos and Labowitz for the Ising model [23].

The basic idea of this algorithm is that at each Monte Carlo step, one process is selected with its corresponding probability and then realized. Because of this approach, the algorithm is not slowed down by unsuccessful attempts. The $k^{th}$ time step of the simulation is as follows:

**Algorithm 2**

Step (1):    Determine all processes *'a'* that possibly could take place with the actual configuration of the system.

Step (2):    Calculate the rate $R_a$ and hence the total rate $Q(C_k)$.

Step (3):    Choose a random number r with uniform distribution in the range [0, $Q(C_k)$).

Step (4):  Find the corresponding event by choosing the first index S for which

$$\sum_{a=1}^{S} R_a(C_k) \geq r.$$

Step (5):  Proceed with the event S leading to a new configuration $C_{k+1}$.

Step (6):  Update the simulation time $t \rightarrow t + \delta\tau$. Update $R_a$ that have changed as a result of event S, update Q(C) and any data structure being used. Where $\delta\tau$ is the time interval between the events and have a Poisson distribution,

$$\delta\tau = \frac{\ln(r_\varepsilon[0, Q(C_k)))}{Q(C_k)} \quad (3.12)$$

This distribution is Poisson because the KMC algorithm simulates Poisson processes.

Considering the dependence of computer time needed on N which is related to system size, first three steps and fifth step are not time consuming but the step four is time consuming [23, 31]. Thus, after generating a random number r, a linear search requires a time O(N) where as if binary search is used, it requires O(logN). Since the growth rules are usually local, the updating in sixth step doesn't have to cost too much computer time, although careful programming is needed. Depending on data structure used, the maximum time needed in step six is O(N) in a linear search.

In order to obtain a faster algorithm, one has to consider the group of events instead of individual events. This is an intermediate scheme of Maksym that groups N rates into n subsets, labeled as $\alpha$ = 1, 2, 3... n performing linear search on the smaller sets. This grouping can be done in different ways such as:

(i)  By forming the groups with the same number of events which allows maximum effectiveness of the algorithm.

(ii)  By forming the groups of same kinds of events which keep the physics clear.

Here, we concentrate explicitly the second case where each group represents a certain process (e.g. deposition of an atom, hopping of an adatom over a specific energy barrier and so on). Each process has a specific rate, $\rho_\alpha$, which is same for all members of the corresponding group.

In the given configuration C, each possible process can be realized by one or more events. Assume that each process α can be realized in $n_α$ ways. The quantity $n_α$ is called the multiplicity of the process. For example, there might be a number of adatoms with identical surroundings which can diffuse, or deposition can take place on a certain number of sites. A relative rate $q_α(C) = n_α \, \rho_α$ is assigned to each kind of process. The total transition rate in a configuration C is now,

$$Q(C) = \sum_{\alpha=1}^{n} n_\alpha(C)\rho_\alpha \tag{3.13}$$

The algorithm in the $k^{th}$ time step of simulation is as follows:

**Algorithm 3**

Step (1): Determine all the processes *'a'* that possibly could take place with the actual configuration C of the system.

Step (2): Compute the overall rate:

$$Q(C_k) = \sum_{\alpha=1}^{n} n_\alpha(C_k)\rho_\alpha$$

Step (3): Select a random number $r_1 \in [0, Q(C_k))$.

Step (4): Search through the list of partial sum $n_\sigma$ until $r_1 \leq n_\sigma = \sum_{\alpha=1}^{\sigma} q_\alpha(C)$ and decide which kind of process will take place.

Step (5): Select an event among the set of events that occur at this rate. Technically, this can be done with the help of a list of conditions for each kind of movement and an integer random number $r_2 \in [1, n_\sigma(C_k)]$. $r_2$ is generated and the corresponding member of the list is selected.

Step (6): Perform the selected movement.

Step (7): Update the configuration C. *i.e.* update $n_α$, $q_α$, $Q(C_k)$ and any data structure being used.

KMC simulations have similar sources of possible errors as that of standard Monte Carlo (for example, defects in pseudo random number generation; finite size effects; and so on). The techniques used to solve these problems are also similar such as periodic boundaries, and so on. It has been found that, in general, thermally excited, time dependent phenomena are not expected to be correctly described by the standard thermodynamical Monte Carlo. The dynamical quantities depend on the choice of transition probabilities, so they are chosen in agreement with physical processes to ensure correct description of the time evolution of the system.

## 3.3 Simulation Models

### 3.3.1 Full- Diffusion Model

Models of epitaxial growth under typical MBE conditions commonly have the following basic features [38]. Atoms are randomly deposited onto an initially flat substrate at a rate of F monolayers per second (ML/s), and they are allowed to diffuse freely until they encounter another atom, a group of atoms, or a defect such as a step. When the flux of incoming particles is turned on, the population of isolated adatoms starts to increase linearly with time. Small clusters (or islands) being to nucleate, when the number of free adatoms becomes sufficiently large. As deposition proceeds, the total number of islands increases and after a while overtakes the number of monomers (isolated adatoms). The islands continue to grow in size and eventually coalesce to form the first complete monolayer. The first stage of the sub monolayer growth, when the monomer density is larger than the total island density, is referred to as the nucleation regime. It is followed by the aggregation regime, which begins when the island density exceeds the monomer density and ends when the island density begins to drop at the initial stages of the coalescence regime. In the aggregation regime the film growth is characterized completely by the dynamical evolution of the island size distribution, island shapes, and locations.

### 3.3.2 The basic model

The KMC simulations in this study are based on the so-called full-diffusion (FD) SOS model with a simple cubic lattice [39,40], which has previously been successfully used in the study of MBE growth. The SOS model permits neither vacancies nor overhangs. The basic model has Arrhenius dynamics and random deposition without additional relaxation. Desorption is forbidden so that the surface coverage is given by $\Theta = Ft$. The dominant mechanism is surface diffusion. The movement of the ad-atoms on the surface can be described as thermally activated hopping processes. Any particle on the surface can move, but only jumps to the nearest neighbor positions are allowed. The hopping rate is defined as the probability of a diffusion jump per unit time which doesn't depend on the bonding energy on the final position (the position after hop). For a surface adatom the hopping rate is characterized by Arrhenius-type expression i.e. $k_0 \exp(-\frac{E}{K_B T})$. Where, E is the energy barrier to hopping, T is the substrate temperature, and $K_B$ is Boltzmann's constant. The prefactor $k_0 = \frac{K_B T}{h} \approx 10^{13} s^{-1}$, is the vibration frequency of a surface ad-atom, where $h$ is Planck's constant [41].

The hopping energy barrier, E, is a sum of two contributions, a site-independent surface energy term $E_S$ and a term given by the number of lateral nearest neighbors, $nE_N$, where $E_N$ is the in-plane bond energy. The total barrier to hopping is thus

$$E = E_S + n\, E_N \tag{3.13}$$

Where n = 0,1,2,3,4 is the number of occupied lateral nearest neighbors at the initial site. In this basic model of diffusion, the adatom lands with equal probability at any of the four neighboring sites.

The above mentioned model can be modified in different ways: both the rules for deposition and the rules for diffusion can be modified. Instead of depositing the arriving particles at purely randomly selected sites, an incorporation mechanism may be included in the model. In this case, a surface site is first selected randomly, and then a search for the site with maximum number of nearest neighbors is carried out within a square of fixed linear size $2R_i+1$ lattice constants centered upon the incidence site [42].

Various modifications have been made in the rates of diffusions. FD model has been modified by introducing interaction with next- nearest neighbors in planes below and above the hopping atom in order to mimic Ehrlic- Schwoebel effect. One can modify the basic model considering the effect of bond energy $E_{NN}$ of next nearest neighbors m= 0, 1, 2.,3, 4.

Recently, the basic model has been modified by including the position dependent strain energy $E_{strain}$ or incorporating the pattern substrate by dividing the lattice into square shaped domains. These modifications have become useful in order to obtain self- assembled nanostructures or quantum dots.

Our modification to FD model is due to anisotropic crystal surface with two non equivalent directions X and Y: the in- channel and cross channel direction respectively. This means adatom can diffuse along X and Y direction with different probability. With these considerations, the energy barrier for the surface diffusion along the X- direction and Y- direction are respectively

$$E_X = E_{S,X} + n_X E_{N,X} + n_Y E_{N,Y} \qquad (3.14)$$

$$E_Y = E_{S,Y} + n_X E_{N,X} + n_Y E_{N,Y} \qquad (3.15)$$

Where the numbers $n_x$, $n_Y$ =0, 1, 2 represent the number of bonds of nearest neighbors in X and Y direction respectively.

## 3.4 Models used and Simulation

We have the simplest model as used by Saito [43] and Heyn [44], in this model, atoms are depositing on a (100) surface of a simple cubic lattice with a solid-on-solid (SOS) restriction where no overhangs and no vacancies are allowed. Atoms are deposited randomly with a deposition rate F monolayers per unit time onto the substrate on which impurities are deposited at fixed separation forming the square. It should be noted that absolute values of the temperatures (T) are not important in the simulation since T scales with E. But if E is kept constant, the temperature plays significant role. The ratio of E to T changes with changing value of T. The temperature is assumed to be low that all deposited atoms stick to the surface and never evaporate back into the ambient gas. An isolated adatom hops to one of its four nearest neighboring sites at a rate *ks* and the surface diffusion constant is given by *Ds*= $k_s a^2$, where *'a'* is the lattice constant. As for the interlayer diffusion, which takes place in case a deposited atom lands on another adatom, no additional energy barrier is assumed for it to diffuse down. On the contrary, atoms at the edge are not allowed to hop up a layer. Therefore, Enrlich-Schwoebel mound formation is suppressed, and the layer-by-layer growth is expected.

When an atom touches to other adatoms or islands, it makes a nearest neighbor bond to form a cluster. At very low temperatures, this cluster will never dissociate again. We assume that if adatoms attach to impurities then adatoms do not dissociate, but if the adatoms are attached forming single bond to each other then dissociation takes place at moderate temperature. This means that dimmers may not be stable at moderate temperature. So island size increased with increasing substrate temperatures by dissociation of small islands. The effect of strain should be more pronounced at higher temperatures when the bigger islands are formed. Also due to higher temperature is should be easier for an atom to detach from the uncomfortable position in the island edge of a more strained island [45]. Thus at high temperature an adatom detachment takes place even from big sized islands. If an edge atom is singly-bonded in the layer, i.e. if it makes only one bond with another adatom, it is possible to diffuse along the island periphery at moderate temperatures. There are two possible diffusion jumps for a singly-bonded atom; an edge diffusion along the straight step edge, and a corner diffusion or corner crossing round the outer corner. The rate of edge diffusion is $k_e$ and that of corner is $k_c$, as is depicted in Fig. 3.3.

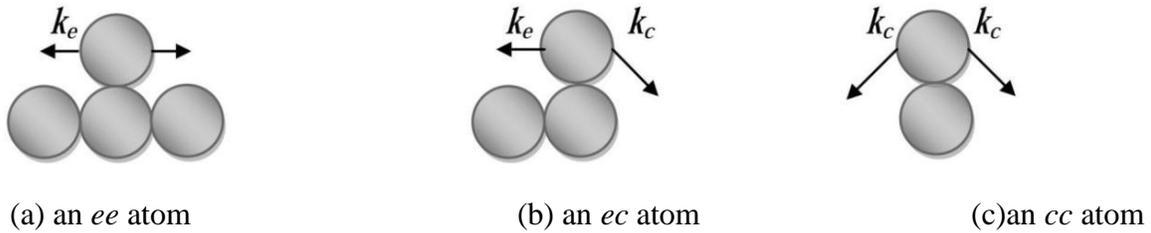

(a) an *ee* atom             (b) an *ec* atom             (c) an *cc* atom

Fig. 3.3: Diffusion process of an adatom from an island with a single in-layer bond.

A singly-bonded atom can be classified as (a) an 'ee' atom on a straight step if both movements are along the step edge, (b) an 'ec' atom at the ridge of a corner if one is the edge diffusion and the other is the corner, and (c) a 'cc' atom on a tip if both movements are across the corner, as shown in Fig 3.3. The edge diffusion constant along the straight step edge is $De = k_e a^2$, and we may define the corner diffusion constant as $Dc = k_c a^2$. Precise values of diffusion constants depend on the energy barriers and the temperature. For edge and corner diffusion there may exist extra energy barriers in addition to that for the surface diffusion, and diffusion constants are probably ordered as $Ds \geq De \geq Dc \geq 0$.

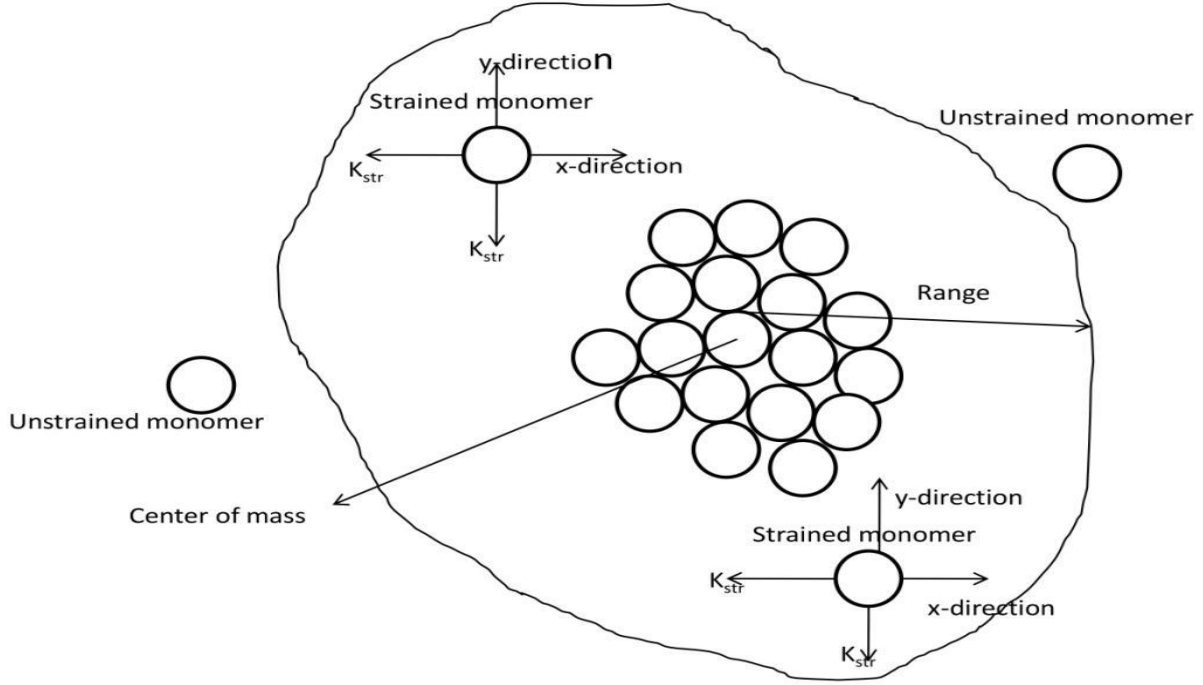

Fig. 3.4: Diffusion of strained monomer

The effect of strain is to lower the barrier which enhanced the detachment of perimeter adatoms from a strained islands[46]. However in our model, simplifications were made by taking the constant value of strain energy equal to 0.028eV which is considered as negative that enhanced an adatom detachment rate from strained islands and form monomer. If a monomer is within the range of bigisland then it is considered as strained monomer. Here the bigisland stands for the island containing more than 10 adatoms. The range of bigisland is calculated by using the concept of center of mass. There are possible diffusion jumps for the strained monomers: two jumps along x-direction and two jumps along y direction. These atoms are influenced by the surface energy and also by the strain energy. This strain energy is considered as negative which enhances the diffusion rate of strained atoms. Thus the diffusion rate of strained monomer is larger than that of monomer. The strained monomer diffusion constant is $D_{str} = k_{str} a^2$. The rate of strained monomer diffusion is kstr as depicted in the Fig. 3.4. The adatom single bonded with the impurity is not allowed for edge and corner diffusions, as shown in the Fig. 3.5 below. This is due to the high binding energy between impurity atom and adatom.

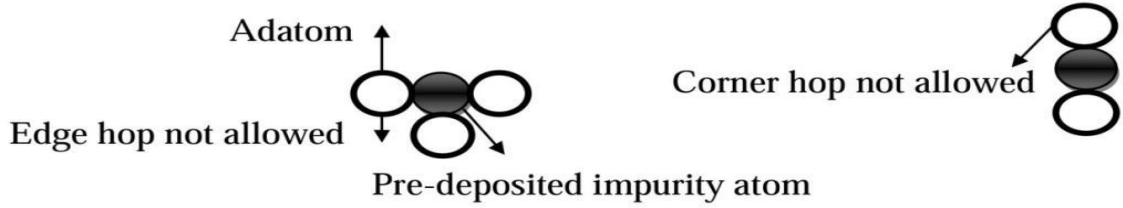

Fig. 3.5: Nature of the bond of adatom with impurity

Simulation starts from substrate surface of a size 300×300 with certain numbers of impurities placed in the fixed separations. Hereafter we take the lattice constant $a=1$. Therefore, the jump rates $k$"s and diffusion constants $D$"s are the same. During the simulation, at some stage, there are N0 isolated adatoms, Nee edge atoms attached to a straight step edge, Nec atoms at the corner, and Ncc atoms at the tip position (as shown in Fig. 3.3) on the surface. These adatoms are sorted out in corresponding lists. The transition probability from this configuration to the next one is given as follows. The rate of the deposition is $q_d=FL^2$, of the surface diffusion $q_s=4N_oDs$, of the edge diffusion $q_e=(2Nee+Nec)De$, of the corner diffusion $q_c=(Nec+2Ncc)Dc$, of the detachment $q_{dt}=(2Ncc+Nec+Nee)Dt$, of the strain-driven detachment $q_{sdt}=(2Ncc+Nee)Dst$ and of the strain surface diffusion$(q_{str})=4NsDstr$. The total probability of a state change in a unit of time is $q_t=q_d + q_s + q_e + q_c + q_{dt} + q_{sdt} + q_{str}$. Thus, within a time interval $dt=1/qt$, one of the events takes place; the deposition $q_ddt$, the surface diffusion $q_sdt$ the edge diffusion with $q_edt$, the corner diffusion with $q_cdt$, the detachment with $q_{dt}dt$, strain-driven detachment $q_{sdt}dt$ and the strain surface diffusion with $q_{str}dt$. For each diffusion and detachment process, the atom to move is picked up randomly from the list. After each state change, adatom lists have to be adjusted, for example, by elimination those adatoms with more than two neighboring bonds from the lists, or add an adatom that happens to have a single bond to the corresponding list, or the adatom that happens to have the no neighbors. After detachment if the atoms are monomers then the atoms are listed in the monomer list and are allowed to reattach to the same or different island. And after the strain-driven detachment of edge atom from island, the newly formed monomer is listed in the strained monomer list , which is generally remains close to that island and re-attaches to it with a high probability but in the rate equation treatment, an adatom that detached from an island becomes available for capture by all islands[47].

## 3.5 Program Description

The programs are coded using the C programming language. Flow chart of main program is shown in Fig 3.6. We have presented brief description of some main functions of the program as follows:

### 3.5.1 Initialization

A matrix of form S[X][Y] is used to represent the substrate. The elements of matrix S[X][Y] represent lattice sites. At first, we initialize the substrate by setting each element of the matrix S[X][Y] to zero (0). After the program has finished each non zero element of the matrix S[X][Y] represents one atom. This is represented by function *initialization.*

### 3.5.2 Impurity deposition

We consider first 100 atoms as impurity atoms and deposited on the substrate in the two dimensional square lattice with separation of 30 in both x and y direction. Impurity atoms are considered as immobile in our simulation during the simulation. This is represented by function *impurity deposition.*

### 3.5.3 Rate Calculation

Each event happens with unique rate. The rates of all events during simulation are calculated within the function *qrate.* The deposition rate is given by L×L×Coverage. Similarly, monomer diffusion rate along x-direction is given by $Dx = D_0 \times \exp(-Ex/K_BT)$. Where, $D_0$ is the vibrational frequency and is the order of $10^{14}$ per sec. The monomer diffusion rate along y direction is given by $D_y = D_A \times D_X$. Here $D_A$ is diffusional anisotropy parameter. If we set $D_A=1$, then it means the diffusion is isotropic. The degree of anisotropy increases with decreasing value of $D_A$. The strained monomer diffusion rate along x-direction is $Dsx = D_0 \times \exp((-Ex+Ec)/K_BT)$ and the strained monomer diffusion rate along y-direction is $Dsx=Dsy$. Similarly the diffusion rate along x-direction is given by $Dnx = D_0 \times \exp(-(Ex+En)/K_BT)$. The detachment rate is calculated as $Dd = D_0 \times \exp(-Ed/K_BT)$. And the strain-driven detachment rate is given by $Dst = D_0 \times \exp((-Ed+Ec)/K_BT)$.

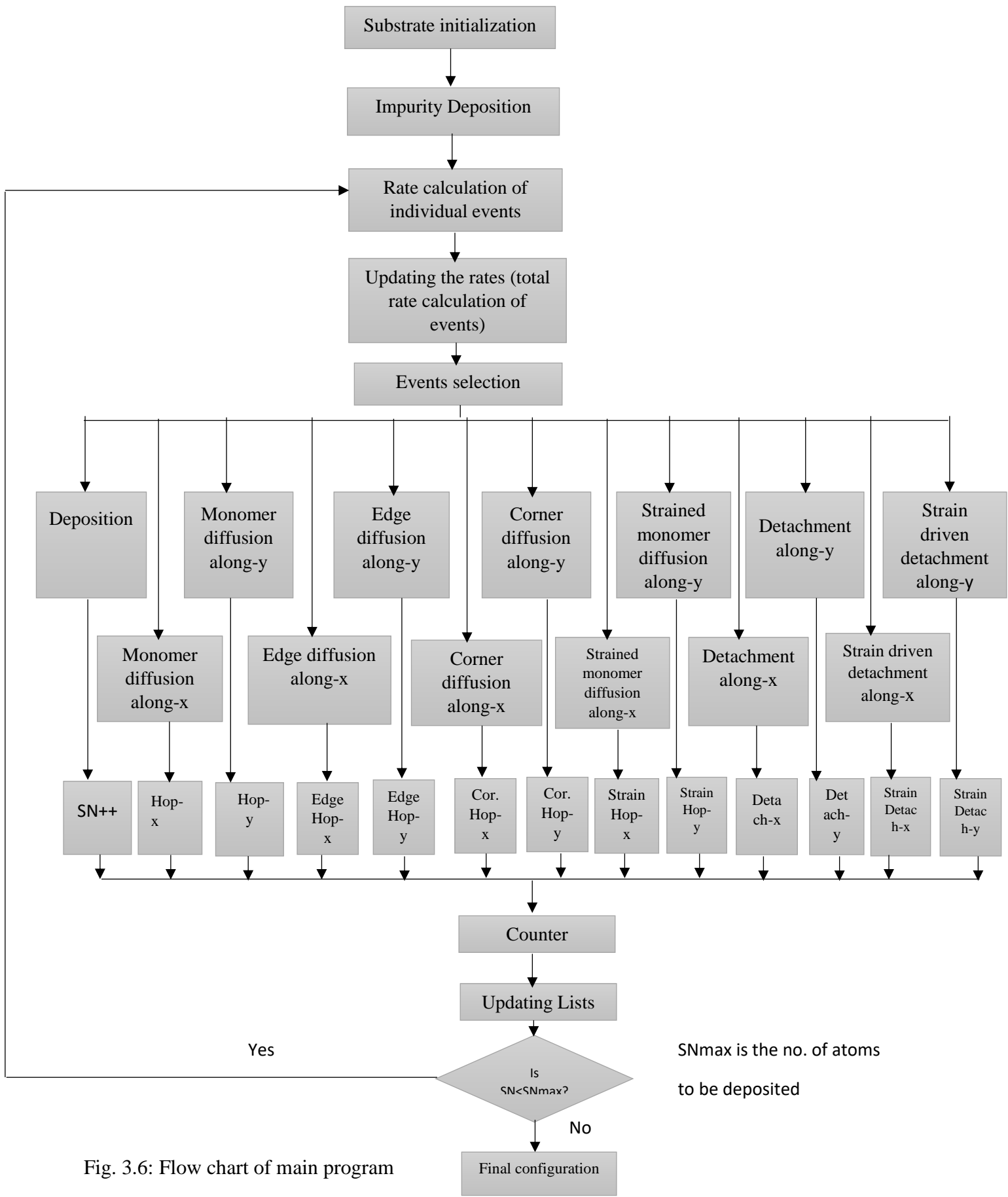

Fig. 3.6: Flow chart of main program

### 3.5.4 Updating the Rates

In this function, we have calculated the total rate for individual event. The total rates of these individual events is summed to get the total rate combining all the events. $total\ rate = \sum q[i]$ where, i=1,2,...,13, *q[i]'s* are the total rates of individual events. The „*update rate*" function is used in the program.

### 3.5.5 Event Selection

This function selects an event which is to be realized. The event, whose rate is high, has a greater probability to be selected. We generate a random number (RANDOM) between 0 and *total rate* and compare it with an integer "$q_k$=$q_k$+q[i]", where, 1≤ i≤ no. of events. The initial value of i is 1 and that of "$q_k$" is equal to the value of the rate of deposition. If $q_k$ ≥ RANDOM, then deposition event is selected otherwise we increase i by 1 and update the value of qk so that its new value is the sum of the rates of deposition and monomer diffusion along x. Again we compare this new value of qk with the random number (RANDOM) i.e. if the condition, $q_k$≥ RANDOM, is satisfied, another event "monomer diffusion along x" would be selected. In this way the selection loop continues until an event is selected. For this function *select event* is used in the program

### 3.5.6 Deposition

The function *deposition* is used to deposit atoms on the substrate. The additional deposited atoms (adatoms) are represented by non-zero numbers whereas vacant lattice sites are represented by 0 on the substrate matrix S[X][Y]. The position where atom is to be deposited is selected randomly. If the selected position is vacant then we keep an atom in that position. Otherwise we select another vacant position randomly. We continue to deposit the atoms until final coverage is reached. After each deposition we increase an integer SN (the total no. of deposited atoms) whose maximum value represents the final coverage known as SNmax.

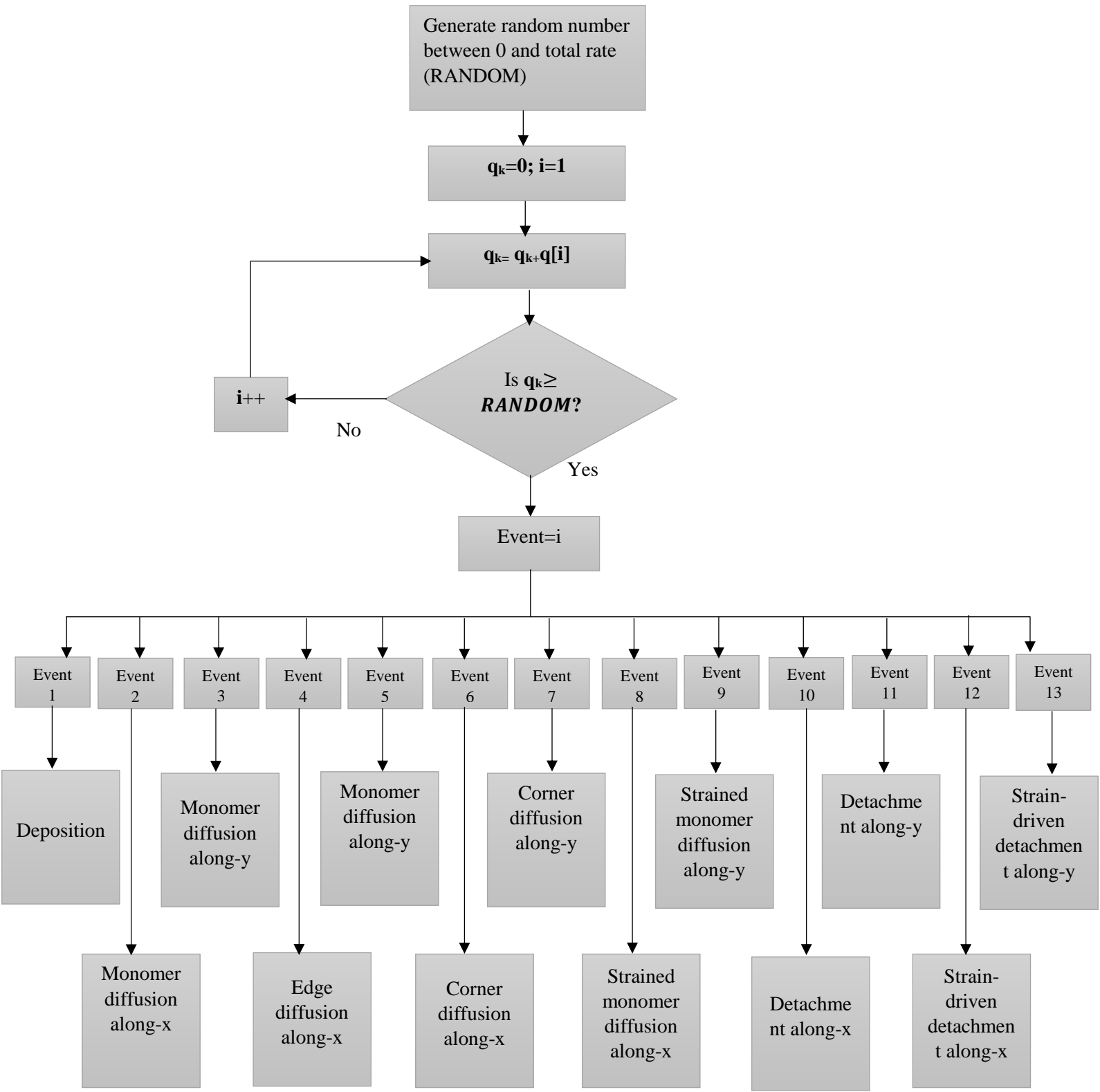

Fig. 3.7: Flow chart for the event selection function

### 3.5.7 Monomer Diffusion

For monomer diffusion we have made two functions, *mono_hoppingx* and *mono_hoppingy* for our convenience. The function, *mono_hoppingx* allows the monomer to diffuse along x-direction whereas, *mono_hoppingy* allows the monomer to diffuse along y-direction. These two functions are quite similar to each other; the only difference is in the direction of hopping. First we pick up an atom randomly from the monomer list. From event selection function the direction of hopping is fixed i.e. either x-direction or y-direction. The new position of hopping is determined randomly.

### 3.5.8 Edge Diffusion

We have made two functions *edgediffusionx* and *edgediffusiony* to simulate this event. The function *edgediffusionx* allows the adatoms to move in x-direction within an island and similarly *edgediffusiony* allows the adatom to move in y-direction. An atom is selected randomly from a list of the edge diffusing atoms. All the edge atoms with single bond are permitted for edge diffusion. But the atom having single bond with the impurity is not allowed to diffusion.

### 3.5.9 Corner Diffusion

We have made two functions *cornerdiffusionx* and *cornerdiffusiony* for the corner rounding event along x-direction and y-direction, respectively. The atom must have single bond for the corner diffusion. But, the atom having single bond with the impurity is not allowed for this event.
If from the select event *cornerdiffusionx* is selected, and Nth atom selected randomly from the list of the corner diffusing atoms and its initial position be N=S[X][Y], the Nth atom must diffuse along x-direction. For x-direction we have two sub functions *corup* and *cordown*. The *corup* sub function moves the Nth atom either in the neighboring position S[X-1][Y+1] or S[X-1][Y-1], provided these positions must be empty and after the diffusion to this new position the Nth atom must lie on the same island as it was before the diffusion. The *corndown* sub function moves the Nth atom either in the neighboring position S[X+1][Y+1] or S[X+1][Y-1], provided these positions must be empty and after the diffusion to this new position the Nth atom must lie on the same island as it was before the diffusion. This can be illustrated more clearly from the Fig. 3.8.

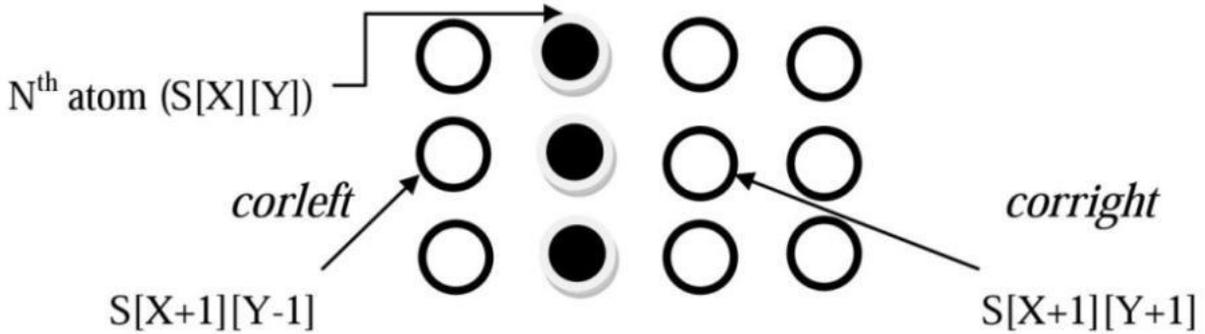

Fig. 3.9: diffusion along y-direction

## 3.5.10 Detachment:

For the detachment of the deposited atoms we defined two functions *detachx* and *detachy* to detach atoms from islands along x-direction and y-direction respectively. An atom is selected randomly from the detach list in which the atoms with single bond are listed, and is allowed to detach randomly selected position in either x-direction or y-direction. Atom which has single bond with impurity, is not allowed to detach.

## 3.5.10.1 Detach-x

If from the select event *detach-x* is selected, $N^{th}$ atom is selected randomly from the list of the single bond atoms for the detachment along x-direction and its position be S[X][Y] then we find the BN value of this atom. If BN=1, this atom is detached to the position S[X-1][Y]. If BN=2, the atom is detached to the position S[X+1][Y]. If BN=4, then we check the bond (BN) of the neighbor atom of position S[X][Y-1], if the neighbor atom has bond value (BN)=9 or 13, atom is detached to the position S[X-1][Y] and if the neighbor atom has bond value (BN)=10 or 14, atom is detached to the position S[X+1][Y]. Similarly, if the neighbor atom has the bond value (BN)=8 or 12, atom is detached randomly either to the position S[X-1][Y] or S[X+1][Y]. If the Nth atom has BN=8, we check the bond value of the neighbor atom of position S[X][Y+1], if this neighbor atom has bond value(BN)=5 or 13, Nth atom is detached to the position S[X-1][Y], if the neighbor atom has bond value(BN)=6 or 14, Nth atom is detached to the position S[X+1][Y]. Similarly, if the neighbor atom has bond value (BN) =4 or 12, Nth atom is detached randomly either to the position S[X-1][Y] or S[X+1][Y]. Finally, the position S[X][Y] is made zero.

### 3.5.10.2 Detach-y

If from the select event *detach-y* is selected, and $N^{th}$ atom is selected randomly from the list of the single bond atoms for the detachment along y-direction and its position be S[X][Y] then we find the BN value of this atom. If BN=4, this atom is detached to the position N=S[X][Y+1]. If BN=8, the atom is detached to the position S[X][Y-1]. If BN=1, we check the bond of the neighbor atom of the position S[X+1][Y], if the neighbor atom has the bond value (BN)=6 or 7, Nth atom is detached to the position S[X][Y+1], if the neighbor atom has the bond value (BN)=10 or 11, Nth atom is detached to the position S[X][Y-1]. Similarly, if the neighbor atom has the bond value (BN) =2 or 3, Nth atom is detached randomly either to the position S[X][Y-1] or S[X][Y-1]. If the Nth atom has BN=2, we check the bond value of the neighbor atom of the position S[X-1][Y], if the neighbor atom has the bond value (BN)=5 or 7, Nth atom is detached to the position S[X1][Y+1], if the neighbor atom has the bond value (BN)=9 or 11 then atom is detached to the position S[X][Y-1]. Similarly, if the neighbor atom has the bond value (BN)=1 or 3, atom Nth is detached randomly either to the position S[X][Y-1] or S[X][Y+1]. The position S[X][Y] is made zero.

### 3.5.11 Strained Monomer Diffusion

For strained monomer diffusion, all the islands consisting of more than 10 adatoms are listed within the function bigisland. The centre of mass of these bigislands are calculated in the function centre of mass. By using the concept of centre of mass, the range of bigislands is determined. Now the distance of adatom from selected event is calculated if it is less than or equal to the range of big island then the adatom is considered to be within the range and it is listed as strained monomer. Here the two functions strainmonomer diffusion x and strain monomer diffusion y are used for the strain monomer hopping event along xdirection and y-direction, respectively. These two functions are quite similar to each other, the only difference is direction of hopping. For this event first of all an adatom is picked up randomly from the list of strained. Here, the two functions strain-monomer *diffusionx* and strain-monomer *diffusiony* are used for the strain monomer hopping event along x-direction and y-direction, respectively. These two functions are quite similar to each other; the only difference is direction of hopping. For this event first of all an adatom is picked up randomly from the list of strained monomer. Then the selected adatom is allowed to move in the randomly selected position either in xdirection or y-direction.

### 3.5.12 Strain-Driven Detachment

The strain considered here arises due to the lattice mismatch between the substrate surface and deposits. Small islands are comparatively less strained than large islands. To obtain the strain field in general one has to use the theory of elasticity, but here for simplicity we will employ a constant value. Strain promotes the detachment of adatoms from the edge of strained islands to facilate the fomation of self-assembled array of nano patterns. We defined two functions *sdetachx* and *sdetachy* for the strain-driven detachment of atom from strained islands along x-direction and y-direction respectively. A singly-bonded atom of island having diffusions along the step edges and across the corners is selected randomly from islands list, and is allowed to strain-driven detach randomly selected position in either x-direction or y-direction. *Atom which has single bond with impurity, is not allowed to have strain-driven detachment.*

### 3.5.12.1 Strain Detach-x

If function *sdetach-x* is selected, $N^{th}$ atom is selected randomly from those islands having diffusions along the step edges and across the corners in x-direction. Nth atom position is S[X][Y] before strain-driven detachment then find the BN value whether to detach in S[X+1][Y] or S[X-1][Y]. If BN=1, this atom is detached to the position S[X-1][Y]. If BN=2, the atom is detached to the position S[X+1][Y]. If BN=4, then we check the bond (BN) of the neighbor atom of position S[X][Y-1], if the neighbor atom has bond value (BN)=9 or 13, atom is detached to the position S[X-1][Y] and if the neighbor atom has bond value (BN)=10 or 14, atom is detached to the position S[X+1][Y]. Similarly, if the neighbor atom has the bond value (BN)=8 or 12, atom is detached randomly either to the position S[X-1][Y] or S[X+1][Y]. If the Nth atom has BN=8, we check the bond value of the neighbor atom of position S[X][Y+1], if this neighbor atom has bond value(BN)=5 or 13, Nth atom is detached to the position S[X-1][Y], if the neighbor atom has bond value(BN)=6 or 14, Nth atom is detached to the position S[X+1][Y]. Similarly, if the neighbor atom has bond value (BN)=4 or 12, Nth atom is detached randomly either to the position S[X-1][Y] or S[X+1][Y]. Finally, the position S[X][Y] is made zero.

### 3.5.12.2 Strain Detach-y

If function *sdetach-y* is selected, $N^{th}$ atom is selected randomly from those islands having diffusions along the step edges and across the corners in y-direction. Nth atom position is S[X][Y] before strain-driven detachment then find the BN value whether to detach in S[X][Y+1] or S[X][Y-1]. If BN=4, this atom is detached to the position N=S[X][Y+1]. If BN=8, the atom is detached to the position S[X][Y-1]. If BN=1, we check the bond of the neighbor atom of the position S[X+1][Y], if the neighbor atom has the bond value (BN)=6 or 7, Nth atom is detached to the position S[X][Y+1], if the neighbor atom has the bond value (BN)=10 or 11, Nth atom is detached to the position S[X][Y-1]. Similarly, if the neighbor atom has the bond value (BN)=2 or 3, Nth atom is detached randomly either to the position S[X][Y-1] or S[X][Y-1]. If the Nth atom has BN=2, we check the bond value of the neighbor atom of the position S[X-1][Y], if the neighbor atom has the bond value (BN)=5 or 7, Nth atom is detached to the position S[X1][Y+1], if the neighbor atom has the bond value (BN)=9 or 11 then atom is detached to the position S[X][Y-1]. Similarly, if the neighbor atom has the bond value (BN)=1 or 3, atom Nth is detached randomly either to the position S[X][Y-1] or S[X][Y+1]. The position S[X][Y] is made zero.

### 3.5.13 Counter

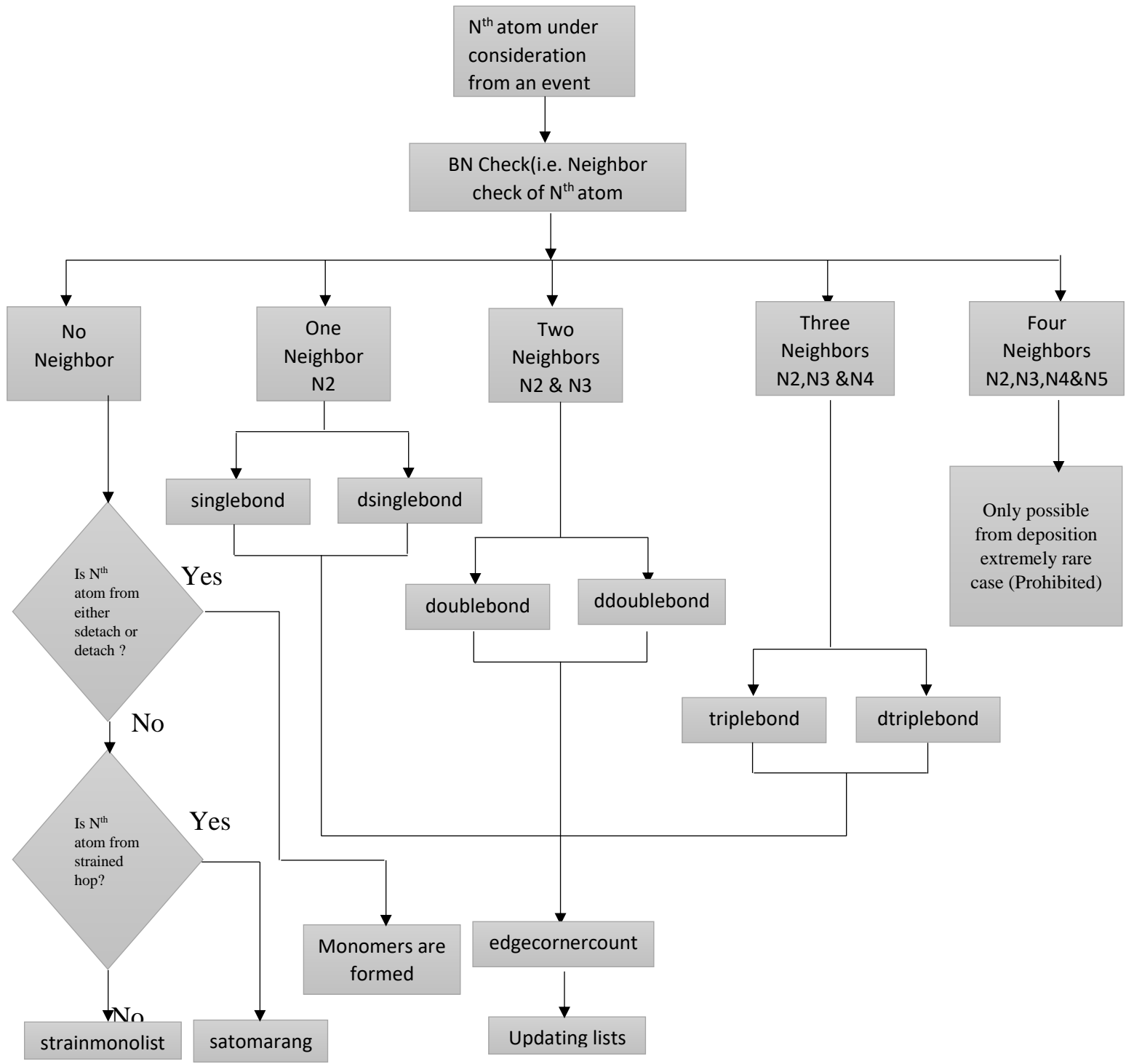

Fig. 3.8: The Program Block Diagram for Counter Function.

In this function, first of all the bond of the considered $N^{th}$ atom is checked after the occurrence of an event to give the address for all the adatoms on the substrate. For this main sub-functions like *Singlebond, Doublebond*, *Triplebond, dsinglebond, ddoublebond* and *dtirplebond* are the defined. The $N^{th}$ atom, which suffers any of the events and finds one or more than one or none of the atoms as nearest neighbors on the substrate, is passed to these functions. If $N^{th}$ atom does not meet any of neighbors but if it is either from strain-driven detachment or from thermal detachment then momomer atom is formed, but if the considered atom is from strained monomer hop then it is passed to the function Satomarng otherwise passed to the function either Strain-monolist. If the $N^{th}$ atom meets a nearest neighbor then it makes single bond with that neighbor. Similarly, if it meets two (or three) nearest neighbors then it makes double (or triple) bonds respectively. If the $N^{th}$ atom is either from strain-driven detachment or from detachment event then it calls the functions either *dsinglebond*, or *ddoublebond* or *dtirplebond.*

### 3.5.13.1 Singlebond and dsinglebond

If $N^{th}$ be the atom from any events except any detachments and it make single bond then program enter sub function singlebond but if $N^{th}$ atom is from either strain-driven detachment or detachment and make single bond then program enter the sub function dsinglebond. If $N^{th}$ be the atom from any events except any detachments and N2 be its neighbor atom then counter calls this subfunction (single bond). The program check the bond (BN) value of the neighbor N2, if the neighbor atom N2 has single bond (i.e BN=1, 2, 4, 8) with $N^{th}$ atom and $N^{th}$ atom is not from the corner diffusion event, a new island is formed. $N^{th}$ and N2 atoms are kept in the new island and the number of the island of size two is increased by one. If N2 is not impurity, monomer number is decreased by 2 but if N2 is impurity then monomer number is decreased by 1. But, if the N2 is in the island and $N^{th}$ is not from the edge diffusion and corner diffusion events, $N^{th}$ atom is kept in that island. The number of the new size island is increased by one and the number of the old size island (i.e. old size=new size-1) is decreased by one, the monomer number is also decreased by one. If $N^{th}$ atom is from edge diffusion or the corner diffusion then the program calls the new function *edgecornercount* (since, no new atom is attached to the island, its function is to recognize the new position of $N^{th}$ atom after the diffusion and assign the new bond value of the $N^{th}$ atom). Finally, to arrange the monomers the program calls the *monolistarng* sub-function to rearrange the monomer numbers. If the $N^{th}$ atom is from either strain-driven detachment event or from detachment event, the program calls the subfunction *dsinglebond,* all the process are same as in the *singlebond.* But, the number of the islands of the size less one than the size of the island from which the atom is detached is increased by one. Finally, it calls the sub-function *islandarrangement* which arranges the island number.

### 3.5.13.2 Doublebond and ddoublebond

If $N^{th}$ atom is from either strain-driven detachment event or detachment event and now forming a double band then it call sub function ddoublebond but if it is from any other event than detachments it call subfunction doublebond. If $N^{th}$ atom be the atom from any events except any detachments and N2 and N3 are neighbor atoms then counter calls this sub-function. The program checks the bond value (BN) of these neighboring atoms. If the neighbor atoms N2 and N3 have single bond with $N^{th}$ atom and $N^{th}$ atom is not from corner diffusion event, a new island is formed. All three atoms are kept in this new island and the number of the new island with size three is increased by one. If N2 and N3 are not the impurities then monomer number is decreased by three. But, if N2 and N3 are the impurities then monomer number is decreased by one. If either N2 or N3 is impurity then monomer number is decreased by two. If $N^{th}$ atom is from the corner diffusion event, the program calls the sub-function *cornercounter*. $N^{th}$ atom is not from the corner diffusion and edge diffusion events and either N2 or N3 is in the island then two atoms are kept in that island and the number of the new sized island is increased by one but the island with size less than two is decreased by one and if either N2 or N3 isn't the impurity, monomer number is decreased by two otherwise it is decreased by one. But, if $N^{th}$ atom is from the edge diffusion but not from the corner diffusion and if either N2 or N3 isn't the impurity, monomer number is decreased by one. If $N^{th}$ atom is not from the corner diffusion and edge diffusion events and if both the neighbors are in the different islands then these two islands are merged and $N^{th}$ atom is kept in that island. But, if $N^{th}$ atom is from the corner diffusion event the program calls sub-function *cornercounter*. Finally, their size and the monomer numbers are arranged. If the $N^{th}$ atom is from either strain-driven detachment event or from detachment event, the program calls the subfunction *ddoublebond,* all the process are same as in the *doublebond.* But, the number of the islands of the size less one than the size of the island from which the atom is detached is increased by one. Finally, it calls the sub-function *islandarrangement* which arranges the island number.

### 3.5.13.3 Triplebond and dtirplebond

If the $N^{th}$ atom is form either strain-driven detachment event or from detachment event, the program calls the sub-function *dtirplebond* and if it is from any other event except any detachments it call sub-function triplebond. If $N^{th}$ be the atom from any events except any detachments and N2, N3 and N4 are neighbor atoms then counter calls this sub-function. The program checks the bond value (BN) of these neighboring atoms. If the neighbor atoms N2, N3 and N4 have single bond with $N^{th}$ atom and $N^{th}$ atom is not from corner diffusion and edge diffusion events, a new island is formed. All four atoms are kept in this new island and the number of the new island with size 4 is increased by one. The number of the island with size less than four is decreased by one. If N2, N3 and N4 are not the impurities, monomer number is decreased by four. If all these neighbor atoms are impurities, monomer numbers is decreased by one. If any one or two of them are impurities, monomer number is decreased by either one or two. But, if the $N^{th}$ atom is from the corner diffusion event then the program call the sub-function *cornercounter*. If any one of the neighbor atoms is in the island and $N^{th}$ atom is not from the corner and edge diffusion events, other three atoms are kept in that island. If remaining two neighbor atoms are not the impurities then monomer numbers are decreased by three. The number of the island with new size is increased by one but the island with size less than three is decreased by one. The monomer number is decreased by three but if the $N^{th}$ atom is from the edge diffusion event then the island with size less than two is decreased by one and monomer number is decreased by two. If any two neighbor atoms are in the different island and $N^{th}$ atom is not from the corner diffusion event then two island are merged and one remaining neighbor atom and $N^{th}$ atoms are kept in that island and corresponding island numbers and monomer number are arranged. But, If two neighbor atoms are in the same island then remaining one atom and $N^{th}$ atom are kept in that island and number of the island and monomer are arranged. But, if $N^{th}$ atom is from the corner diffusion then program calls the sub-function *cornercounter*. If all the three neighbor atoms are in the different island then all these island are merged to form the single island and $N^{th}$ atom is kept in that island and corresponding island and monomer are arranged. But, if all the neighbor atoms are in the same island, $N^{th}$ atom is kept in that island and number of the island and monomer number are arranged. Finally, program calls the sub-function *monolistarg*. If the $N^{th}$ atom is from either strain-driven detachment event or from detachment event, the program calls the subfunction *dtirplebond,* all the process are same as in the *tirplebond.*

But, the number of the islands of the size less one than the size of the island from which the atom is detached is increased by one. Finally, it calls the sub-function *islandarrangement* which arranges the island number.

### 3.5.13.4 Strainmonolist and Satomarng

If considered atom is from strained monomer hop then it is passed to the function satomarng. If the consider atom is from monomer hopping then it is passed to the function strainmonolist. If $N^{th}$ atom from any events except any detachment events does not meet any neighbor then the atom is passed to the function either in strainmonolist or satomarng. If we consider atom is from strained monomer hop then it is passed to the function satomarng. In this function the atom is checked wheather the atom is still in range of any bigisland or not. If it is still within the range of bigisland, it is kept in strained monomer list as earlier otherwise it is removed from strained monomer list. If the consider atom is from monomer hopping then it is passed to the function strainmonolist. In this function atom is checked whether it is within the range of bigisland or not. If it is in range of any big island, it is inserted in the list of strained monomer and remove from monomer list.

# CHAPTER 4

## RESULTS AND DISCUSSION

In this dissertation, we have studied the effect of different strain parameter and temperature of the substrate when atoms deposited on the substrate surface with inclusion of island dependent strain-driven detachment process and strain monomer diffusion on the island morphology, average island size, average island density and relative width distribution during the sub monolayer epitaxial growth. KMC simulation is the tool for our study on the island formation on different substrates with isotropic energy parameters i.e. $Esx=Esy$, where $Esx$ and $Esy$ are the energy parameters for the surface diffusion along X and Y directions, respectively. In this simulation, we first deposit a regular mesh of impurity atom forming a two dimensional square lattice on the substrate. The distance between the pre-deposited impurity atoms is $30a$, where "a" is inter-atomic distance. The number of impurity pre-deposited on the substrate is 100. The pre-deposited impurities are considered as immobile and serve as the nucleation sites. For the simulation, following energy parameters were taken. As the previous studies, the surface energy for diffusion as $Esx=Esy=0.4eV$, the binding energy to nearest neighbor $En=0.1eV$, detachment energy $Ed=0.6eV$ are taken but the value of the strain energy $Ec= \alpha*N$ eV (N is number of atoms in the islands) which lowers the detachment energy barrier. The simulations are performed for the systems of lattice size 300*300 with the coverages 0.01ML, 0.03ML, 0.05ML, 0.08ML and 0.10ML and strain parameters $10^{-6}eV$, $10^{-5}eV$, $10^{-4}eV$, $10^{-3}eV$ and $10^{-2}eV$. After average is taken from 10 different configurations obtained from the simulation parameter, we calculate the average island size, average island density, relative width distribution and island size distribution. The optimal set of growth conditions at which the most uniform islands formed are obtained. We have also determined the value of temperature at which the islands are most equally sized and regularly arranged for different value of the strain parameter.

## 4.1 Effect on Island Morphology

Fig.4.1.1 shows the island morphologies for the five different coverage values 0.01ML, 0.03ML, 0.05ML, 0.08ML, and 0.3ML and five different strain parameter values $10^{-2}$eV, $10^{-3}$eV, $10^{-4}$eV, $10^{-5}$eV and $10^{-6}$eV and constant flux 0.3ML/s at temperature 260K. In all morphologies white spot represents the ad-atoms deposited and the black spot in the middle of the islands represents the impurity atoms. From the figure we find that the average island size increases with increase in the value of coverage for strain parameter $10^{-2}$eV. Similar result is found for all the values of strain parameter ($10^{-6}$eV- $10^{-2}$eV). We find that the at all coverage the islands are nucleated at other site as well as at impurities site. We also find that for all values of coverage large and small islands are formed but at high coverage value the island sizes are widely varied as compared to low coverage value.

From figures 4.1.1 (e), (j), (o), (t) and (y) for the fix value of coverage 0.1ML along the vertical column, we find that as strain parameter increases the average island size also increases and variation in island sizes are decreased. Similar results found for all coverage value 0.01ML, 0.03ML, 0.05ML and 0.08ML. Fig.4.1.2 and Fig. 4.1.3 show the island morphologies for the five different coverage values 0.01ML, 0.03ML, 0.05ML, 0.08ML, and 0.3ML and five different strain parameter values $10^{-2}$eV, $10^{-3}$eV, $10^{-4}$eV, $10^{-5}$eV and $10^{-6}$eV and constant flux 0.3ML/s at temperature 280K and 300K respectively. From the figure we find that as in the case of T=260K, the average island size increases with increase in the value of coverage for all the values of strain parameter. We find that the at all coverage the islands are nucleated at other site as well as at impurities site but appear less as compared to T=260K. From the Fig. 4.1.2 and Fig.4.1.3 (e), (j), (o), (t) and (y) for the fix value of coverage 0.1ML along the vertical column we find that as strain parameter increases the average island size also increases and variation in island sizes are decreased. Similar results found for all coverage value 0.01ML, 0.03ML, 0.05ML and 0.08ML.

T=260K

Covg=0.01ML  Covg=0.03ML  Covg=0.05ML  Covg=0.08ML  Covg=0.10ML

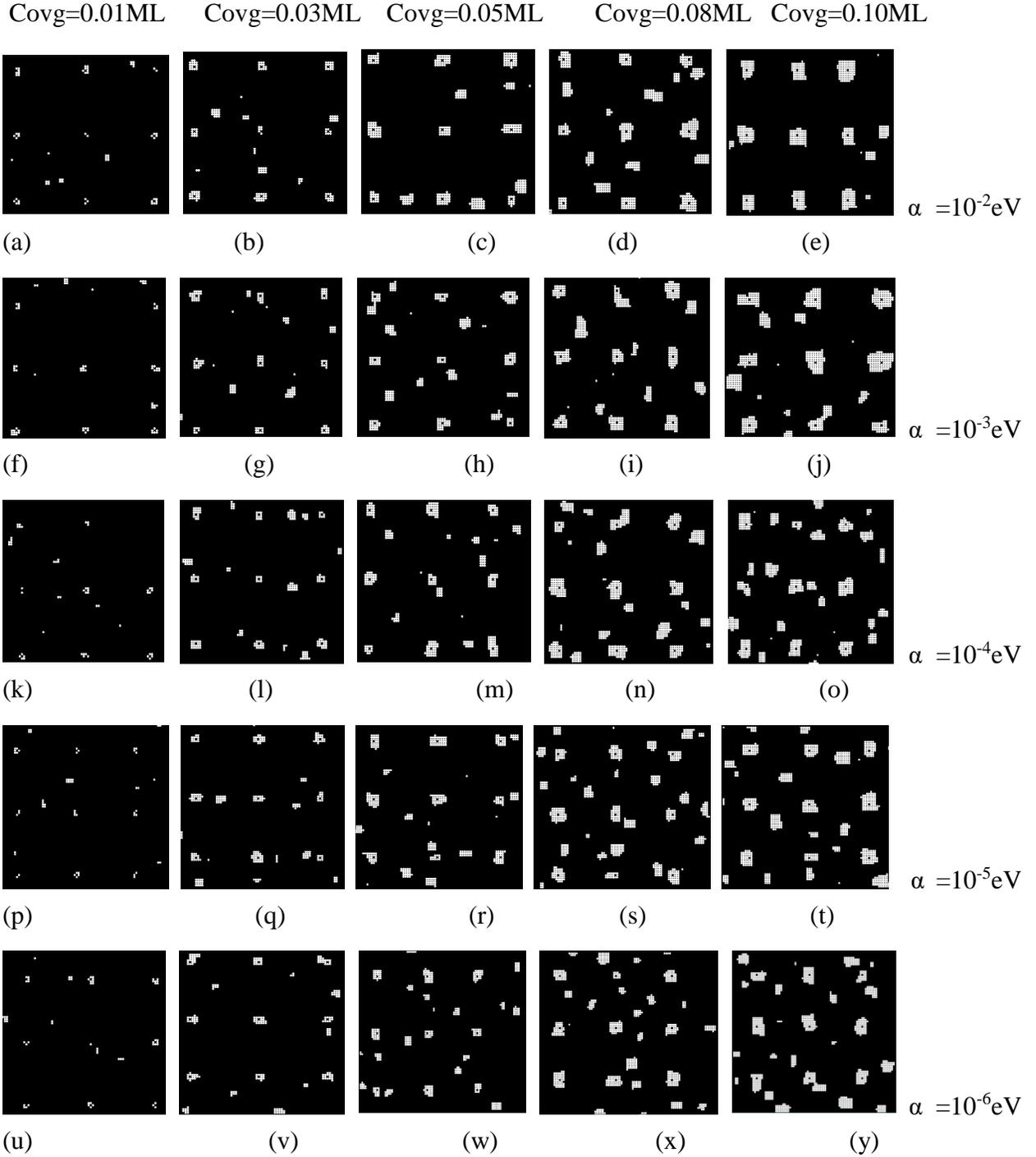

$\alpha = 10^{-2}$eV

(a)　　　(b)　　　(c)　　　(d)　　　(e)

$\alpha = 10^{-3}$eV

(f)　　　(g)　　　(h)　　　(i)　　　(j)

$\alpha = 10^{-4}$eV

(k)　　　(l)　　　(m)　　　(n)　　　(o)

$\alpha = 10^{-5}$eV

(p)　　　(q)　　　(r)　　　(s)　　　(t)

$\alpha = 10^{-6}$eV

(u)　　　(v)　　　(w)　　　(x)　　　(y)

Fig. 4.1.1 Island morphology for various coverage and strain parameter. The simulations parameters are system size L=300, flux=0.3ML/s impurity separation=30a, number of impurity=100, $E_s$=0.4eV, $E_n$=0.1eV, $E_d$=0.6eV & T=260K.

T=280K

| Covg=0.01 ML | Covg=0.03 ML | Covg=0.05 ML | Covg=0.08 ML | Covg=0.10 ML | |
|---|---|---|---|---|---|
| (a) | (b) | (c) | (d) | (e) | $\alpha=10^{-2}$eV |
| (f) | (g) | (h) | (i) | (j) | $\alpha=10^{-3}$eV |
| (k) | (l) | (m) | (n) | (o) | $\alpha=10^{-4}$eV |
| (p) | (q) | (r) | (s) | (t) | $\alpha=10^{-5}$eV |
| (u) | (v) | (w) | (x) | (y) | $\alpha=10^{-6}$eV |

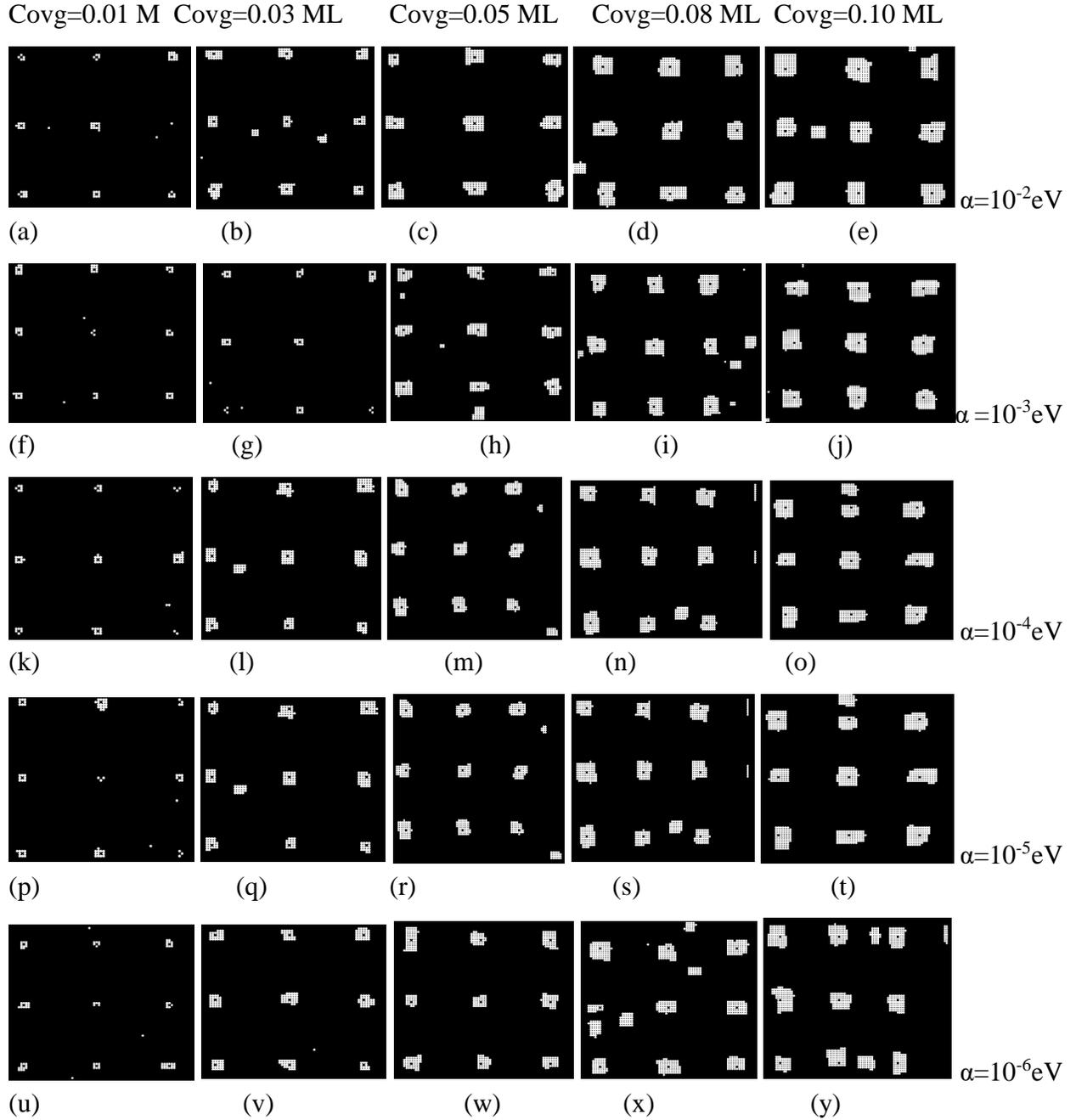

Fig. 4.1.2: Island morphology for various coverage and strain parameter. The simulations parameters are system size L=300, flux=0.3ML/s impurity separation=30a, number of impurity=100, Es=0.4eV, En=0.1eV, Ed=0.6eV & T=280K.

T=300K

Covg=0.01ML   Covg=0.03ML   Covg=0.05ML   Covg=0.08ML   Covg=0.10ML

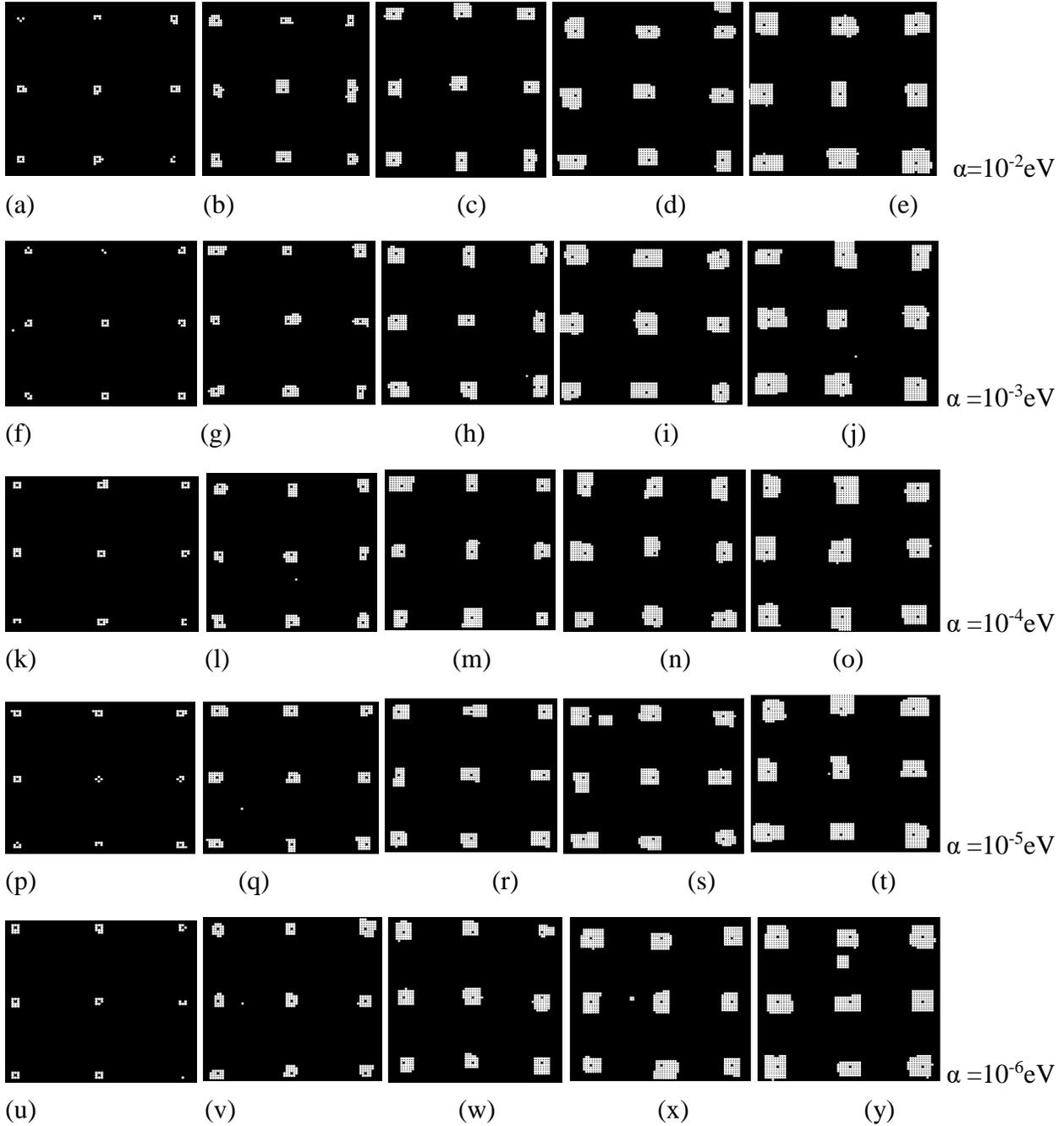

Fig. 4.1.3 Island morphology for various coverage and strain parameter. The simulations parameters are system size L=300, flux=0.3ML/s impurity separation=30a, number of impurity=100, $E_s$=0.4eV, $E_n$=0.1eV, $E_d$=0.6eV & T=300K.

.

From Above morphologies it is clear that, strain parameter plays crucial role to growth of patterned surface. At low strain parameter ($\alpha = 10^{-6}$eV), the nucleation of islands at position other than impurities can see frequently (Fig 4.1.1, (u)-(y)). But as the strain parameter increases, the number of the islands formed at position other than impurities is decreases (Fig 4.1.1, (a)-(e)). So, higher strain parameter plays crucial role to form more patterned form of the islands. In this dissertation, we have performed simulation with flux 0.3ML/s. Earlier dissertation [48] concluded that optimal flux is nearly around 0.025ML/s and in high flux the islands were formed randomly. But, even at high flux, in our case we observed this format of more patterned islands.

At low coverage, the islands are small in size. This is due to smaller number of ad-atoms in the substrate. But in the higher coverage, there is large number of ad-atoms on the substrate which causes to form big island at the position of the impurities [14]. This is because, strain monomer diffusion and strain driven detachment play crucial role to detach ad-atoms from islands. Those detached ad-atoms have great probability to join the islands formed at impurities [48]. That's why at higher strain parameter the growth seems more order in the form as shown in Fig 4.1.2((e),(j),(o),(t),(y)). For high temperature (300K) the island formed at the position of the impurities and growth seems more regular for all values of strain parameter. This is because strain parameter as well as temperature of the substrate plays important role to growth in more ordered in size as well as position [14].

## 4.2. Effect on the Average Island size

Here, we have calculated the average island size using the relation [47]:

$$<s> = \frac{\Sigma SNs}{\Sigma Ns} \qquad S>1, \qquad (4.1)$$

Where, "S" is the number of adatoms on the island and "Ns" is the areal density of islands composed of "S" ad-atoms. Figure 4.2.1 shows average island size as a function of coverage at flux 0.3ML/s and temperature 260K. From the figure it is clearly seen that the average island size increases continuously with increase in a coverage value for all strain parameter. It is seen from the Fig. 4.2.1 that at low coverage 0.01ML the value of the average island size is almost same for all values of the of the strain parameter. But at high coverage value 0.10ML, the value of average island size continuously increases with increase in the strain parameter. This is due to fact that strain plays crucial role to detach ad-atoms from the islands. The strain lowers the detachment energy barrier and hence the ejection of edge ad-atom enhances from large islands. Such detached ad-atom has great probability to capture by islands formed at the position of the impurities [49]. Which results the increase in the average island size. This is more clearly seen in Fig.4.2.2. From Fig. we find that at low coverage (0.01ML), <s> increases very slowly with increase in strain parameter but at high coverage value, the value of average island size continuously increases with increase in the strain parameter. This is due to fact at low coverage there are small number of atoms in the substrate which causes to distribution of only small islands thus value of <s> will be low. Also in low coverage island sizes are small thus probability of formation of island with size > 10 will be low. Thus effect of strain will be low. At high coverage bigger island forms and effect of strain parameters are more pronounced.

Fig.4.2.3 and Fig. 4.2.5 show the average island size as a function of coverage at flux 0.3ML/s and temperature 280K and 300K respectively. From the figure it is clearly seen that at both temperature the average island size increases continuously with increase in a coverage value for all strain parameter. It is seen from Fig. 4.2.3 and Fig. 4.2.5 that at low coverage 0.01ML the value of the average island size is almost same for all values of the of the strain parameter. At high coverage value 0.10ML, the value of average island size increases slowly with increase in the strain parameter as compare with at 260K. This is more clearly seen in Fig.4.2.4 and Fig 4.2.6. From the Fig. we find that at low coverage (0.01ML) <s> increases very slowly with increase in strain parameter, but, even at high coverage value the value of average island size increases slowly with increase in the strain parameter as compare to temperature 260K. This is due to fact that at high temperature the diffusion rate is more as compare to low temperature (260K) due to which probability of atom to reach other island have high which causes to increase in average island size. This is due to fact that strain plays crucial role to detach ad-atoms from the islands, such detached ad-atom has great probability to capture by islands formed at the position of the impurities [49]. The strain lowers the detachment energy barrier and hence the ejection of edge ad-atom enhances form large islands. Which results the increase in the average island size. But in low coverage there are small number of atoms in the substrate which causes to distribution of only small islands thus value of <s> will be low. Also in low coverage island sizes are small thus probability of formation of island with size > 10 will be low.

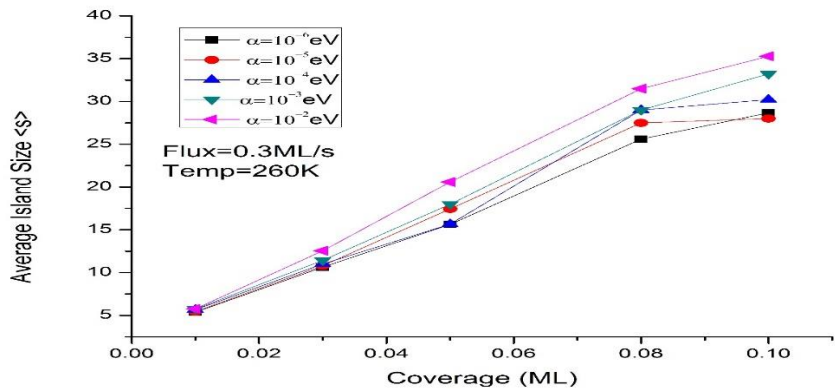

Fig. 4.2.1: Graph of average island size vs. coverage at temperature 260K for different value of strain parameter. The simulation parameters are L=300, coverage=0.1ML, impurity separation=30a where "a" is inter-atomic distance, number of impurity=100, $E_s$=0.4eV, $E_n$=0.1eV, $E_d$=0.6eV, Flux=0.3ML/s and T=260K.

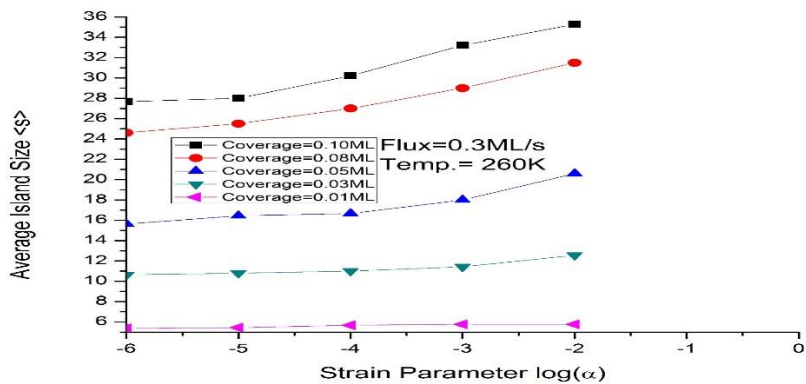

Fig. 4.2.2: Graph of average island size vs. strain at temperature 260K for different value of strain parameter. The simulation parameters are L=300, impurity separation=30a where "a" is inter-atomic distance, number of impurity=100, $E_s$=0.4eV, $E_n$=0.1eV, $E_d$=0.6eV, Flux=0.3ML/s and T=260K.

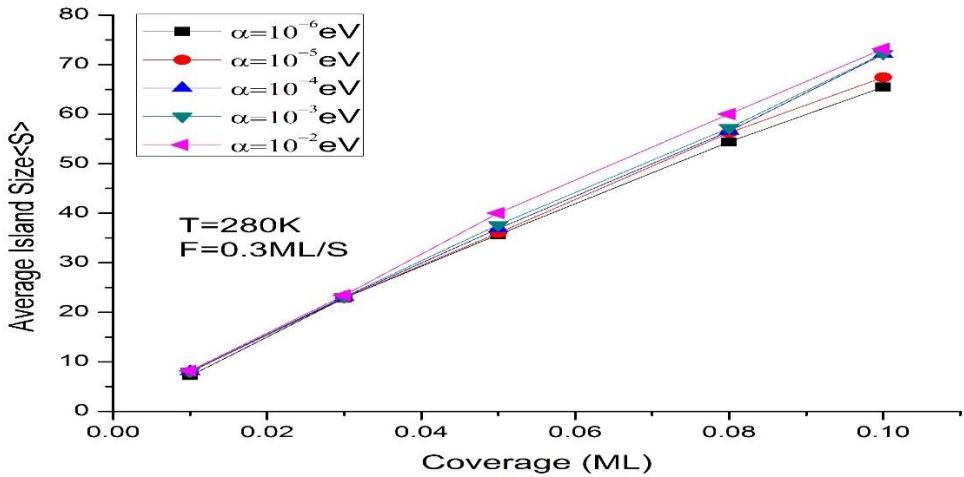

Fig. 4.2.3: Graph of average island size vs. coverage at temperature 280K for different value of strain parameter. The simulation parameters are L=300, coverage=0.1ML, impurity separation=30a where "a" is inter-atomic distance, number of impurity=100, Es=0.4eV, En=0.1eV, Ed=0.6eV, Flux=0.3ML/s and T=280K.

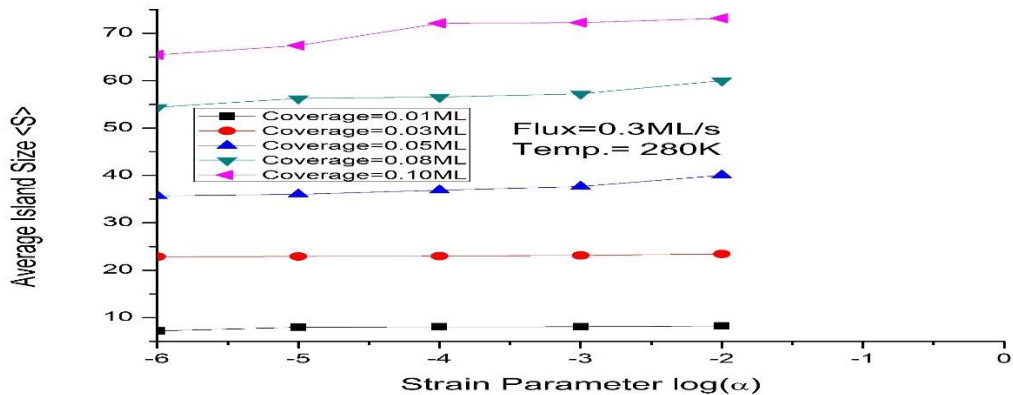

Fig. 4.2.4: Graph of average island size vs. strain at temperature 280K for different value of strain parameter. The simulation parameters are L=300, impurity separation=30a where "a" is inter-atomic distance, number of impurity=100, Es=0.4eV, En=0.1eV, Ed=0.6eV, Flux=0.3ML/s and T=280K.

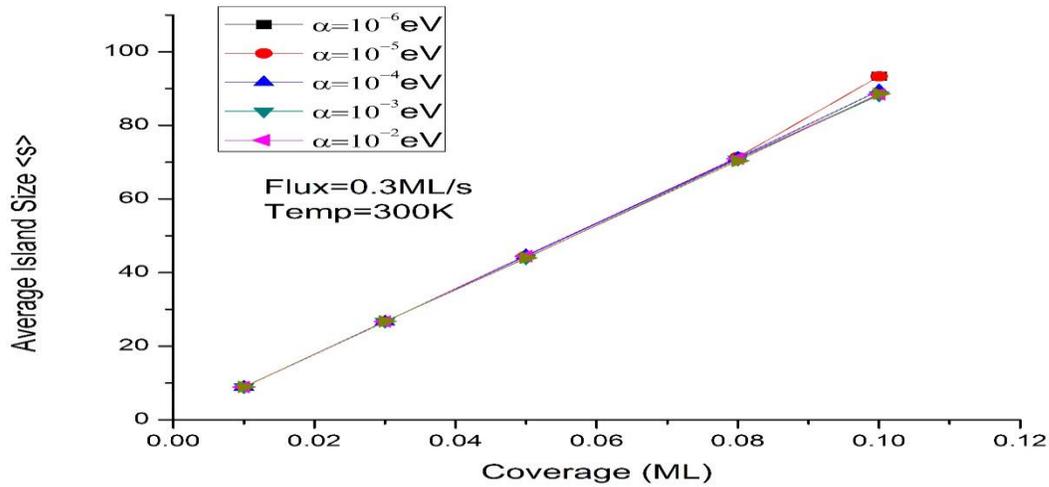

Fig. 4.2.5: Graph of average island size vs. coverage at temperature 300K for different value of strain parameter. The simulation parameters are L=300, coverage=0.1ML, impurity separation=30a where "a" is inter-atomic distance, number of impurity=100, Es=0.4eV, En=0.1eV, Ed=0.6eV, Flux=0.3ML/s and T=300K.

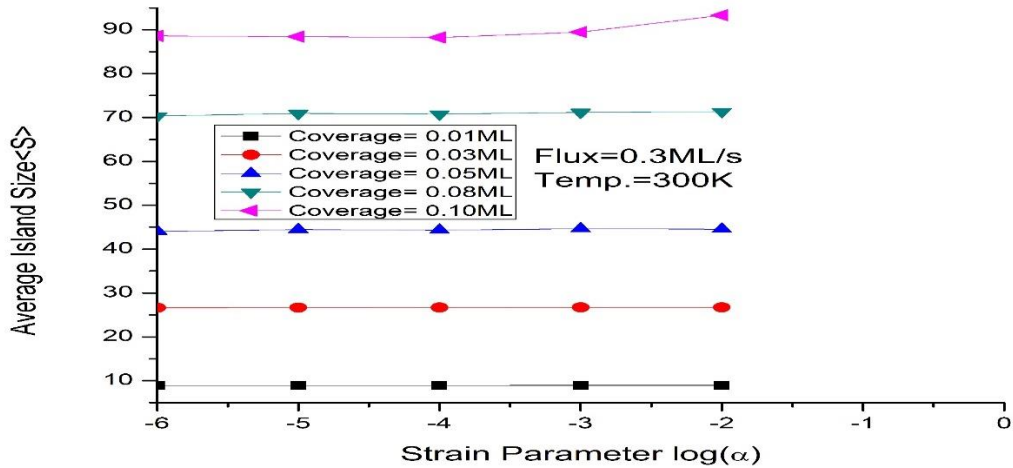

Fig. 4.2.6: Graph of average island size vs. strain at temperature 300K for different value of strain parameter. The simulation parameters are L=300, impurity separation=30a where "a" is inter-atomic distance, number of impurity=100, Es=0.4eV, En=0.1eV, Ed=0.6eV, Flux=0.3ML/s and T=300K.

## 4.3 Effect on the Average Island Density

Here, we have calculated the average island density using the relation:

$$\rho = \frac{\text{Average island number (<N>)}}{\text{Lattice size (L)} \times \text{Lattice size (L)}} \qquad (4.2)$$

Figure 4.3.1 shows that the average island density as a function of coverage at flux 0.3ML/s and temperature 260K. From the figure it is clearly seen that the average island density increases continuously with increase in a coverage value for all strain parameter. It is seen from the Fig. 4.3.1 that at low coverage 0.01ML the value of the average island density is almost same for all value of the strain parameter. But at higher coverage 0.10ML the average island density increases with decrease in the strain parameter. The strain lowers the detachment energy barrier and hence the ejection of edge ad-atom enhances form large islands. This is due to fact that strain plays crucial role to detach ad-atoms from the islands, such detached ad-atom has great probability to capture by islands formed at the position of the impurities. Which results the increase in the average island size and hence decreases the average island density. This is more clearly seen in Fig.4.3.2. From the figure we find that at low coverage (0.01ML), the average island density decreases very slowly with increase in strain parameter but at higher coverage (0.10ML), the value of average island density decreases more rapidly with increase in the strain parameter as compare to low coverage. This is due to fact that at low coverage (0.01ML) there are small number of atoms in the substrate which causes to distribution of only small islands thus value of <s> will be low and average island density is also low. But also at low temperature 260K the diffusion rate is low due to which probability of an atom to reach other island is low which causes the decrease in average island size and hence results the increase in average island density.

Figure 4.3.3 and Fig. 4.3.5 show that the average island density as a function of coverage at flux 0.3ML/s and temperature 280K, 300K respectively. From the figure it is clearly seen that the average island density increases continuously with increase in a coverage value for all strain parameter. It is seen from the Fig. 4.3.3 and Fig. 4.3.5 that at low coverage 0.01ML as well as at high coverage the value of the average island density increases with decrease in the strain parameter very slowly.

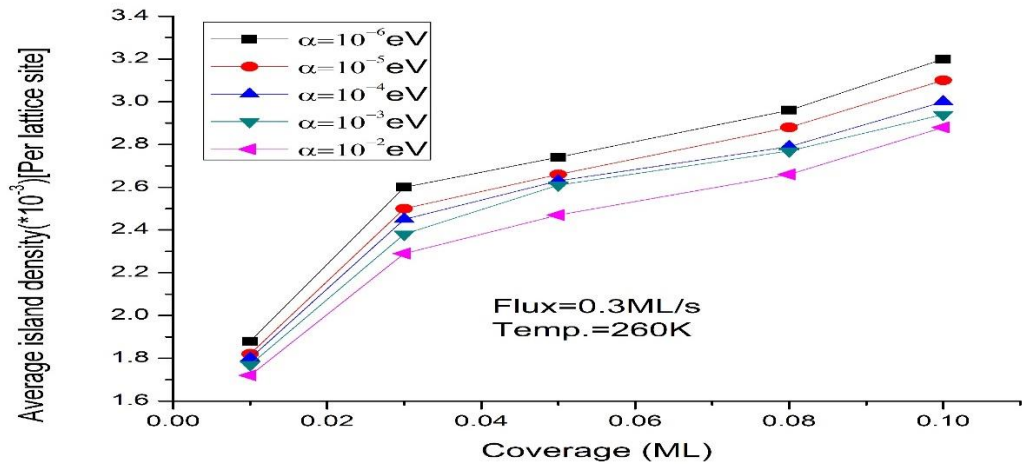

Fig. 4.3.1: Graph of the average island density vs. coverage at temperature 260K for strain parameter 0.01eV, 0.001eV, 0.0001eV, 0.00001eV and 0.000001eV including of strain monomer diffusion and strain driven detachment. The simulation parameters are L=300, Flux=0.3ML/s, impurity separation=30a where "a" is inter-atomic distance, number of impurity=100, $E_s$=0.4eV, $E_n$=0.1eV, $E_d$=0.6eV and T=260K.

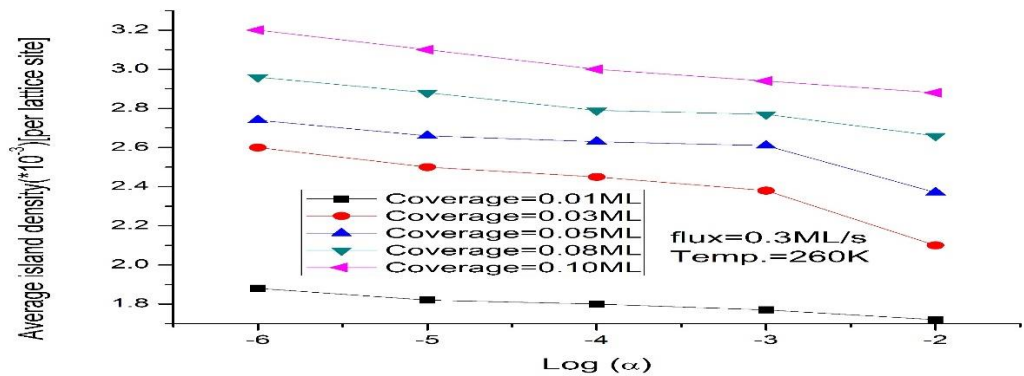

Fig. 4.3.2: Graph of the average island density vs. log (α) at temperature 260K for coverage 0.01ML, 0.03ML, 0.05ML, 0.08ML and 0.10ML including of strain monomer diffusion and strain driven detachment. The simulation parameters are L=300, Flux=0.3ML/s, impurity separation=30a where "a" is inter-atomic distance, number of impurity=100, $E_s$=0.4eV, $E_n$=0.1eV, $E_d$=0.6eV and T=260K.

This is more clearly seen in Fig.4.3.4 and Fig.4.3.6. From the figure we find that at low coverage, the average island density decreases very slowly with increase in strain parameter. This is due to fact that at low strain parameter the atoms are less mobile which lead to the decrease in average island size and hence results the increase in average island density. But at high strain parameter, the atoms are highly mobile due to decrease in binding energy between atoms in islands cause to decrease in average island size and hence increase in average island density. A comparative study as variation of average island density with strain parameter at different temperature (Fig.4.3.2, Fig.4.3.4 and Fig.4.3.6) shows that, as temperature increases curve becomes less step in variation in the average island density. It is because at high temperature the island forms only at impurity sites which are fixed. These island number density will tends to remain constant, due to this the average island density remains almost same.

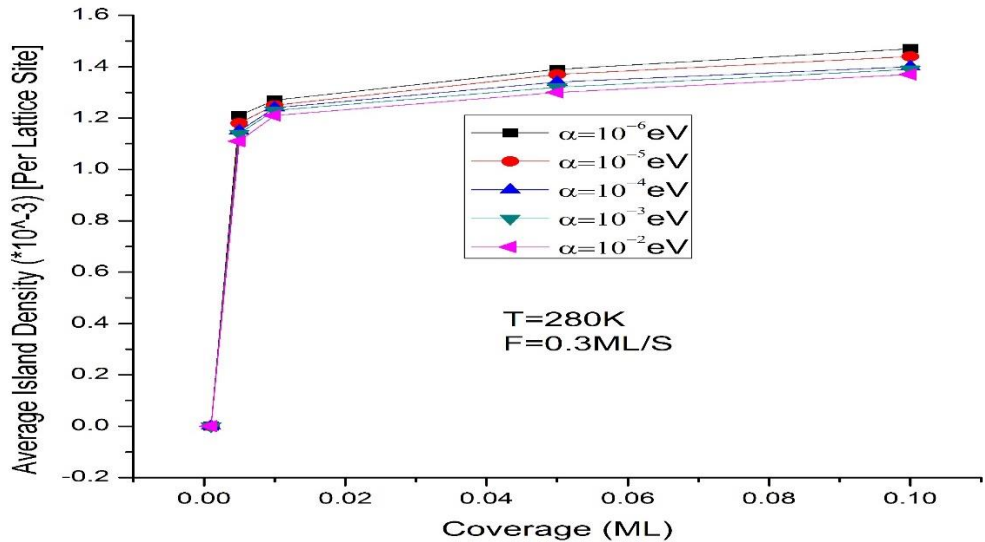

Fig. 4.3.3: Graph of the average island density vs. coverage at temperature 280K for strain parameter 0.01eV, 0.001eV, 0.0001eV, 0.00001eV and 0.000001eV including of strain monomer diffusion and strain driven detachment. The simulation parameters are L=300, Flux=0.3ML/s, impurity separation=30a where "a" is inter-atomic distance, number of impurity=100, $E_s$=0.4eV, $E_n$=0.1eV, $E_d$=0.6eV and T=280K.

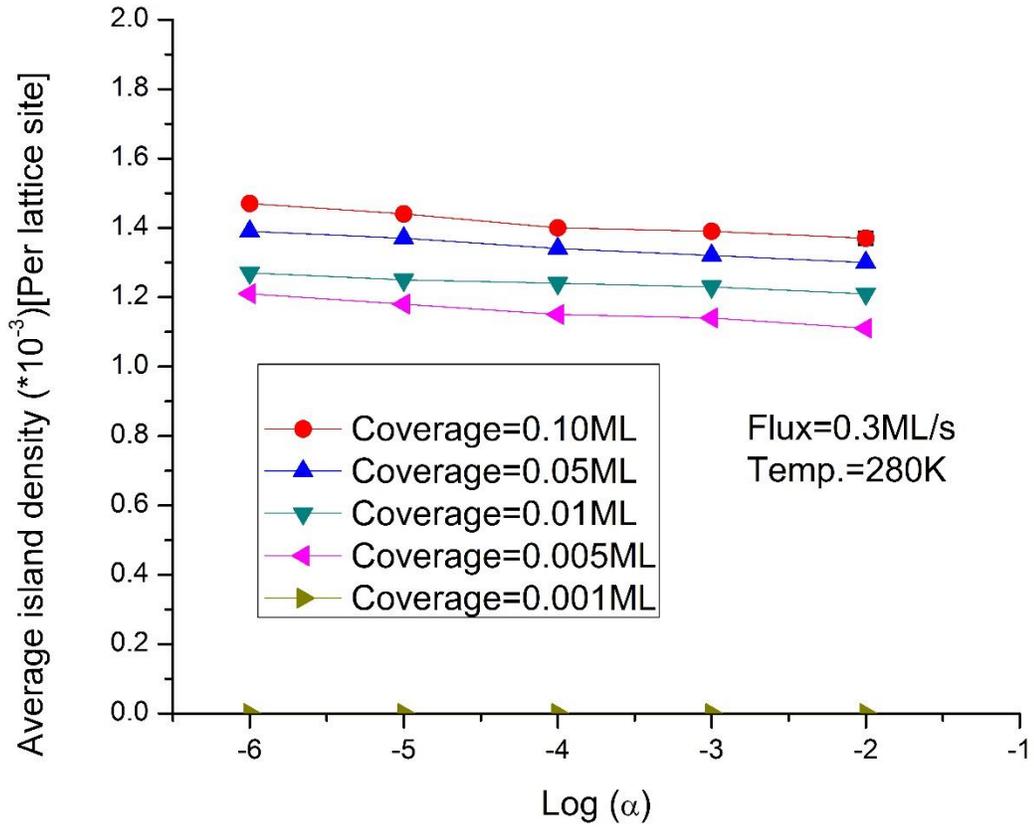

Fig. 4.3.4: Graph of the average island density vs. Log (α) at temperature 280K for coverage 0.01ML, 0.03ML, 0.05ML, 0.08ML and 0.10ML including of strain monomer diffusion and strain driven detachment. The simulation parameters are L=300, Flux=0.3ML/s, impurity separation=30a where "a" is inter-atomic distance, number of impurity=100, $E_s$=0.4eV, $E_n$=0.1eV, $E_d$=0.6eV and T=280K.

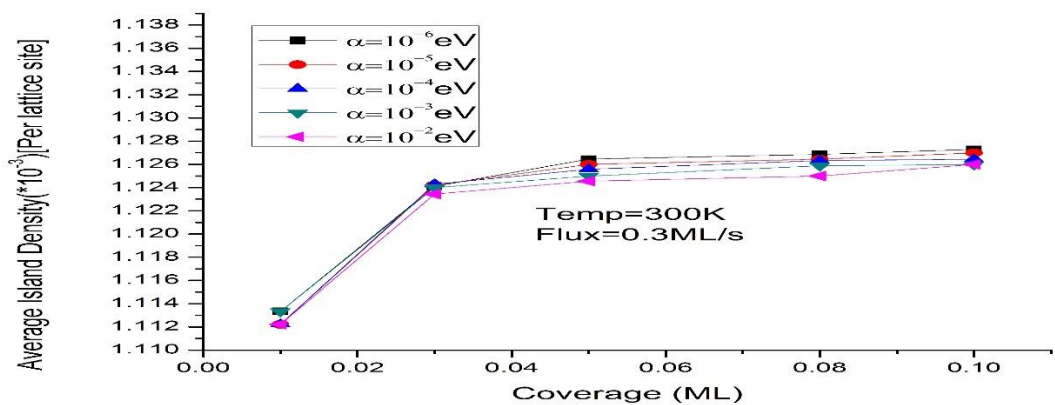

Fig. 4.3.5: Graph of the average island density vs. coverage at temperature 300K for coverage 0.01ML, 0.03ML, 0.05ML, 0.08ML and 0.10ML including of strain monomer diffusion and strain driven detachment. The simulation parameters are L=300, Flux=0.3ML/s, impurity separation=30a where "a" is inter-atomic distance, number of impurity=100, $E_s$=0.4eV, $E_n$=0.1eV, $E_d$=0.6eV and T=300K.

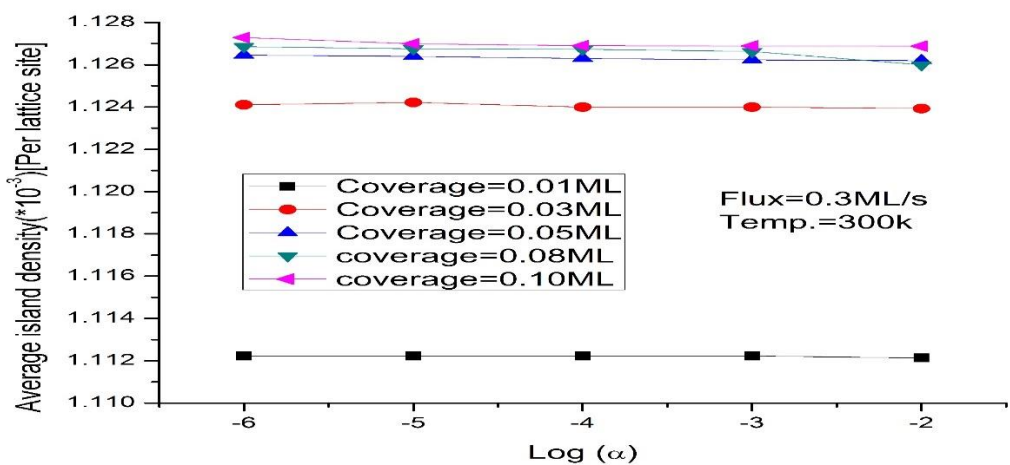

Fig. 4.3.6: Graph of the average island density vs. log(α) at temperature 300K for coverage 0.01ML, 0.03ML, 0.05ML, 0.08ML and 0.10ML including of strain monomer diffusion and strain driven detachment. The simulation parameters are L=300, Flux=0.3ML/s, impurity separation=30a where "a" is inter-atomic distance, number of impurity=100, $E_s$=0.4eV, $E_n$=0.1eV, $E_d$=0.6eV and T=300K.

## 4.4 Effect on the Relative Width Distribution (w)

The relative width distribution (w) at different flux values are calculated by using the relation [16]: $w = \hat{W}/<s>$, where $\hat{W} = <s^2>-<s>^2$ is the width of the size distribution and $<...>$ denotes ensemble average. The smaller $\hat{W}$ is, the smaller fluctuations are in the island sizes and better the size distribution for application. Figs. 4.4.1, 4.4.3 and 4.4.5 shows the plot for relative width distribution vs coverage for different strain parameter α at different fixed temperature 260K, 280K and 300K. From the figure it is seen that the relative width distribution increases continuously with the increase in the coverage value for the all value of strain parameter. At all temperature studied, the highest value of the relative width distribution appears at temperature 260K and lowest seems at 300K. The relative width distribution seems smaller for small value of coverage for all value of strain parameter and three different temperature. But at higher coverage value the relative width distribution is observed to vary with different value of strain parameter at temperature 260K and 280K. But, at 300K, the relative width distribution seems almost same for same value of coverage even for different value of the strain parameter. These are more clearly shows figures. Figs. 4.4.2, 4.4.4 and 4.4.6 shows relative width distribution as a function of strain parameter at three different temperature 260K, 280K and 300K respectively. From figures we find that at low temperature and low coverage the relative width distribution doesn't vary with strain parameter, but at high coverage the relative width distribution decreases with increase in the strain parameter. This is because high strain parameter plays crucial role to growth of islands at high temperature and large coverage. At low coverage sizes of islands are small and number of islands >10 will be small too and therefore strain parameter doesn't play any role on relative width distribution. At large coverage values beyond 0.05ML and at high temperature 280K and 300K and higher strain parameter, the coalescence and aggregation of islands takes place due to the thermal detachment of very small islands atoms other than nucleating at the impurities which results in the formation of larger islands giving the smaller value of relative width distribution. We find that at high coverage value relative width distribution decrease with in strain parameter. It is because greater the island size the single bonded atoms will be more strained and will have high detachment probability. This leads to lower the formation of big islands which leads to formation of more equal sized islands and hence decreases in relative width distribution.

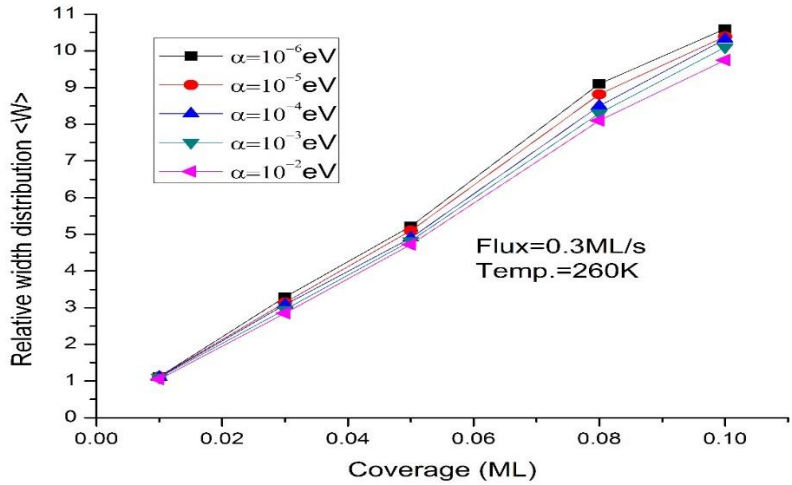

Fig. 4.4.1: Graph of the relative width distribution vs. coverage at temperature 260K, for deposition on pattern impurity with strain-driven detachment and inclusion of strain monomer diffusion for five different value of strain parameter. The simulation parameters are L=300, flux=0.3ML/s, impurity separation=30a where "a" is inter-atomic distance, number of impurity=100, $E_s$=0.4eV, $E_n$=0.1eV, $E_d$=0.6eV and T=260K.

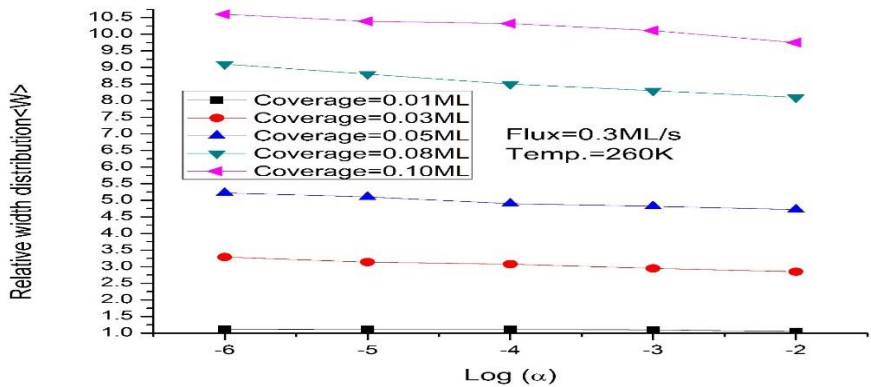

Fig. 4.4.2: Graph of the relative width distribution vs. Strain parameter for different value of coverage, for deposition on pattern impurity with strain-driven detachment and inclusion of strain monomer diffusion for five different value of strain parameter. The simulation parameters are L=300, flux=0.3ML/s, impurity separation=30a where "a" is inter-atomic distance, number of impurity=100, $E_s$=0.4eV, $E_n$=0.1eV, $E_d$=0.6eV and T=260K.

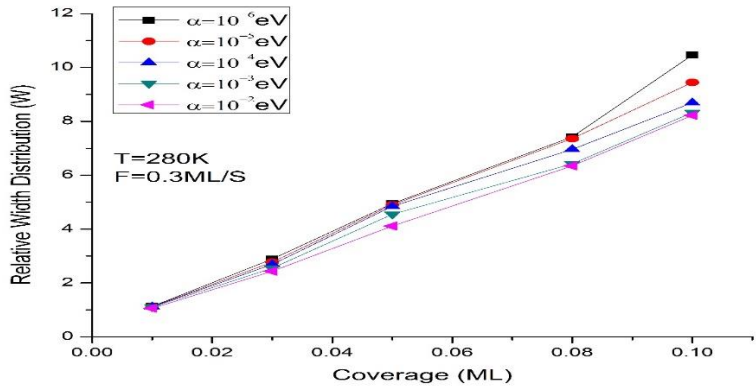

Fig. 4.4.3: Graph of the relative width distribution vs. coverage at temperature 280K, for deposition on pattern impurity with strain-driven detachment and inclusion of strain monomer diffusion for five different value of strain parameter. The simulation parameters are L=300, flux=0.3ML/s, impurity separation=30a where "a" is inter-atomic distance, number of impurity=100, Es=0.4eV, En=0.1eV, Ed=0.6eV and T=280K.

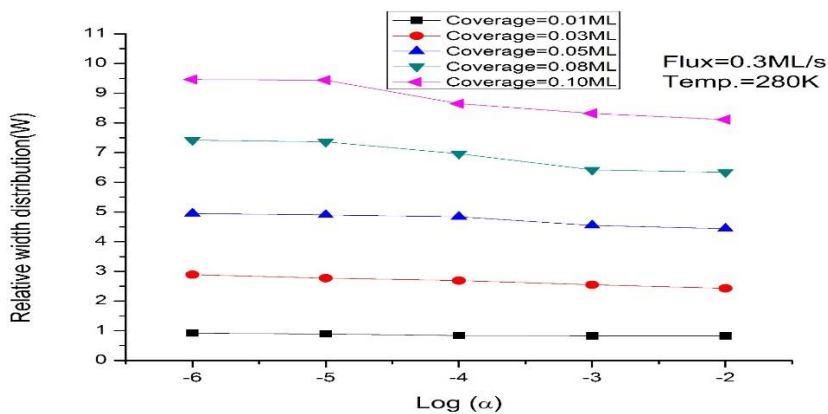

Fig. 4.4.4: Graph of the relative width distribution vs. Strain parameter for different value of coverage, for deposition on pattern impurity with strain-driven detachment and inclusion of strain monomer diffusion for five different value of strain parameter. The simulation parameters are L=300, flux=0.3ML/s, impurity separation=30a where "a" is inter-atomic distance, number of impurity=100, Es=0.4eV, En=0.1eV, Ed=0.6eV and T=280K.

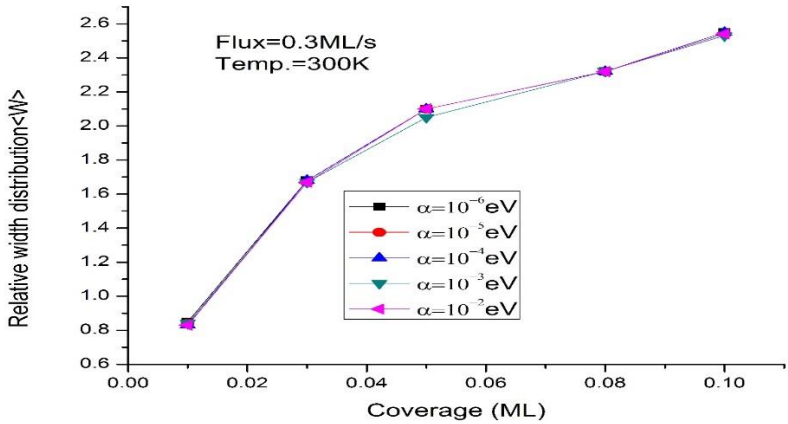

Fig. 4.4.5: Graph of the relative width distribution vs. coverage at temperature 300K, for deposition on pattern impurity with strain-driven detachment and inclusion of strain monomer diffusion for five different value of strain parameter. The simulation parameters are L=300, flux=0.3ML/s, impurity separation=30a where "a" is inter-atomic distance, number of impurity=100, $E_s$=0.4eV, $E_n$=0.1eV, $E_d$=0.6eV and T=300K.

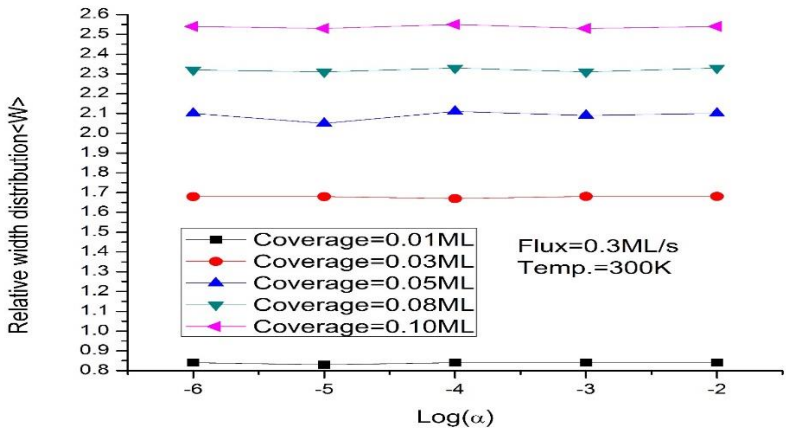

Fig. 4.4.6: Graph of the relative width distribution vs. Strain parameter for different value of coverage, for deposition on pattern impurity with strain-driven detachment and inclusion of strain monomer diffusion for five different value of strain parameter. The simulation parameters are L=300, flux=0.3ML/s, impurity separation=30a where "a" is inter-atomic distance, number of impurity=100, $E_s$=0.4eV, $E_n$=0.1eV, $E_d$=0.6eV and T=300K.

## 4.5 Effects on the Island Size Distribution

We have studied the effect of strain parameter on the island size distribution for the coverage values 0.01ML, 0.03ML, 0.05ML, 0.08ML and 0.10ML at three different temperature. We compared the results of different strain parameter. Fig. 4.5.1 shows the island size distribution as a function of island size at strain parameter $10^{-2}$eV and temperature 260K. From figure we find that at low coverage the width of the curve is small and peak is high therefore islands are observed more in order in size and less value of relative width distribution. But with an increase in coverage the width of the curve increases and peak decreases. This implies that the islands are deviated more from their average island size and increased in the value of relative width distribution.

Fig 4.5.2 depicts the graph between island size distribution as a function of the coverage for deposition of atom on patterned impurity surface including of strain monomer diffusion and strain driven detachment. From the figure we find that peak of the curve decreases and the width of the curve increases with an increase in value of the coverage and peak position also shifted towards the higher value of island size. The decrease in peak and the increase in the width implies that the value of the relative width distribution is more for high coverage than compare to the low coverage. We have also studied the effect of strain parameter in the relative width distribution at 280K. From Fig 4.5.2-Fig 4.5.4 show the island size distribution at different coverage for three different strain parameter. From the figures it is seen that as the coverage increases the peak is decreasing and width is increasing. But at the high strain parameter the peak is less decreasing and width is less increasing than compare to the same curve at low value of α. It implies that the high in value of strain parameter causes to grow in size in more ordered form. This is because as the strain causes to detach ad-atom from island formed and those atoms have high chance to attach on the islands formed at the position of the impurities. Which results the increase in size of the island formed at the position of the impurities. This leads the more ordered in the size and hence decrease in the relative width distribution. So high strain parameter causes to growth of islands more same in size and order giving the consistency in relative width distribution.

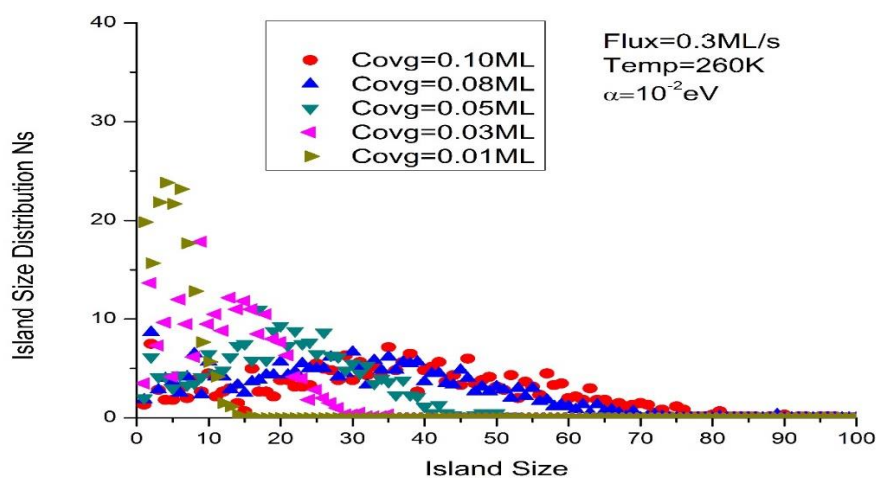

Fig. 4.5.1: Graph of the island size distribution(Ns) vs. island size(S) for flux 0.3 ML/s at T=260K for pattern impurity with strain-driven detachment and inclusion of strain monomer diffusion with strain parameter 0.01eV. The simulation parameters are L=300, flux=0.3ML/s, impurity separation=30a where "a" is atomic distance, number of impurity=100, $E_s$=0.4eV, $E_n$=0.1eV, $E_d$=0.6eV, α =0.01eV.

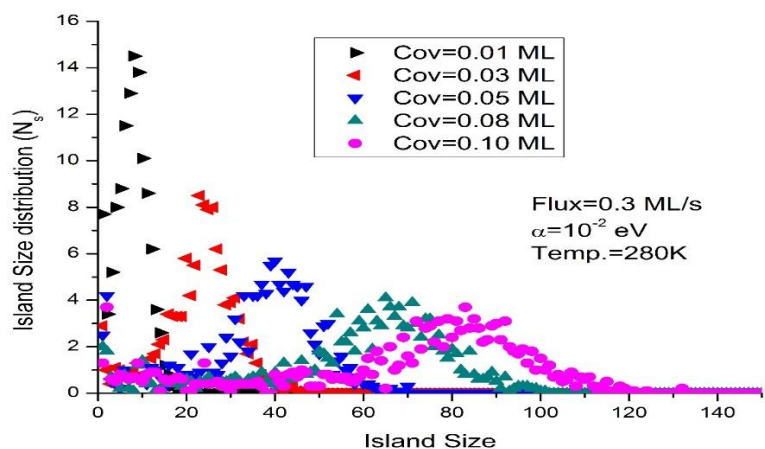

Fig. 4.5.2: Graph of the island size distribution(Ns) vs. island size(S) for flux 0.3 ML/s at T=280K for pattern impurity with strain-driven detachment and inclusion of strain monomer diffusion with strain parameter 0.000001eV. The simulation parameters are L=300, flux=0.3ML/s, impurity separation=30a where "a" is atomic distance, number of impurity=100, $E_s$=0.4eV, $E_n$=0.1eV, $E_d$=0.6eV, α =0.01eV.

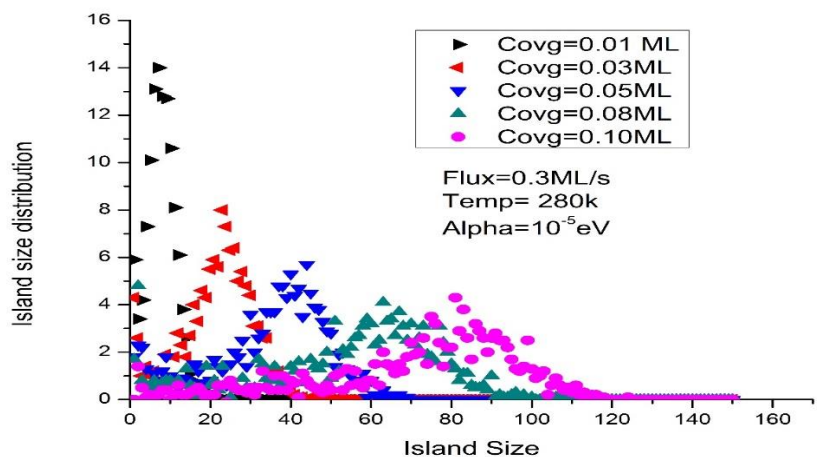

Fig. 4.5.3: Graph of the island size distribution(Ns) vs. island size(S) for flux 0.3 ML/s at T=280K for pattern impurity with strain-driven detachment and inclusion of strain monomer diffusion with strain parameter 0.0001eV. The simulation parameters are L=300, flux=0.3ML/s, impurity separation=30a where "a" is atomic distance, number of impurity=100, Es=0.4eV, En=0.1eV, Ed=0.6eV, α =0.00001eV.

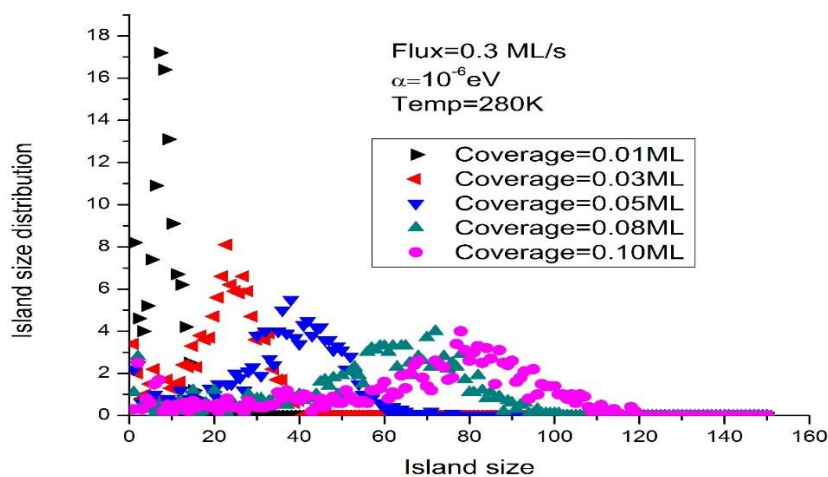

Fig. 4.5.4: Graph of the island size distribution(Ns) vs. island size(S) for flux 0.3 ML/s at T=280K for pattern impurity with strain-driven detachment and inclusion of strain monomer diffusion with strain parameter 0.001eV. The simulation parameters are L=300, flux=0.3ML/s, impurity separation=30a where "a" is atomic distance, number of impurity=100, Es=0.4eV, En=0.1eV, Ed=0.6eV, α =0.000001eV.

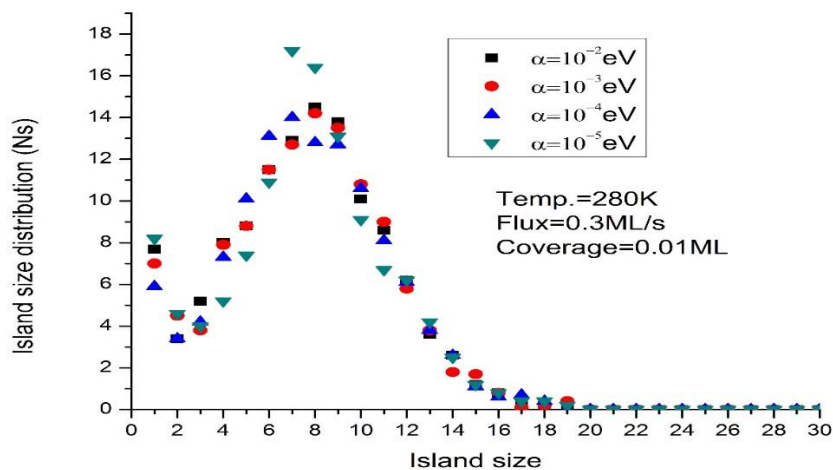

Fig. 4.5.5: Graph of the island size distribution(Ns) vs. island size(S) for flux 0.3 ML/s at T=280K for pattern impurity with strain-driven detachment and inclusion of strain monomer diffusion. The simulation parameters are L=300, flux=0.3ML/s, impurity separation=30a where "a" is atomic distance, number of impurity=100, $E_s$=0.4eV, $E_n$=0.1eV, $E_d$=0.6eV, Coverage=0.01ML.

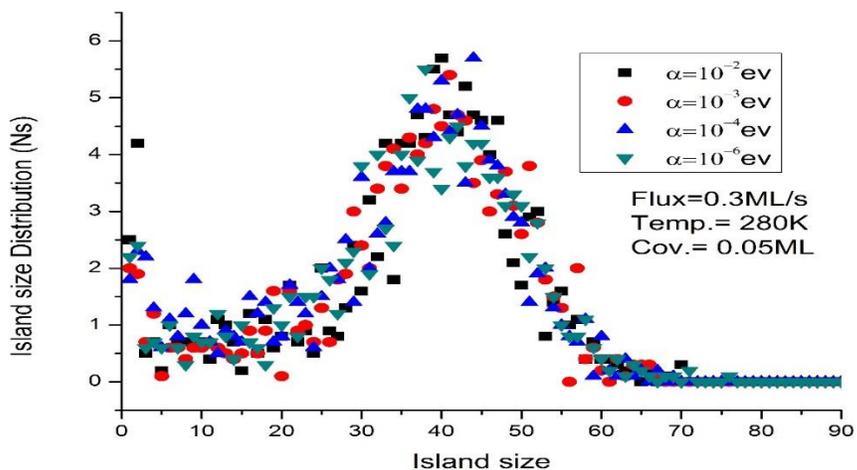

Fig. 4.5.6: Graph of the island size distribution (Ns) vs. island size(S) for flux 0.3 ML/s at T=280K for pattern impurity with strain-driven detachment and inclusion of strain monomer diffusion. The simulation parameters are L=300, flux=0.3ML/s, impurity separation=30a where "a" is atomic distance, number of impurity=100, $E_s$=0.4eV, $E_n$=0.1eV, $E_d$=0.6eV, Coverage=0.05ML.

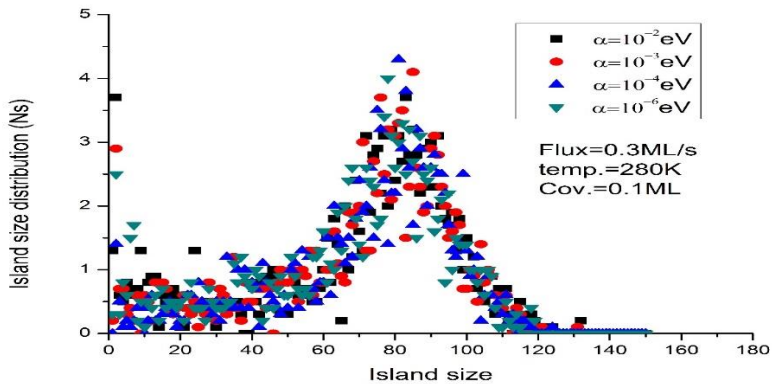

Fig. 4.5.7: Graph of the island size distribution (Ns) vs. island size(S) for flux 0.3 ML/s at T=280K for pattern impurity with strain-driven detachment and inclusion of strain monomer diffusion. The simulation parameters are L=300, flux=0.3ML/s, impurity separation=30a where "a" is atomic distance, number of impurity=100, Es=0.4eV, En=0.1eV, Ed=0.6eV, Coverage=0.10ML.

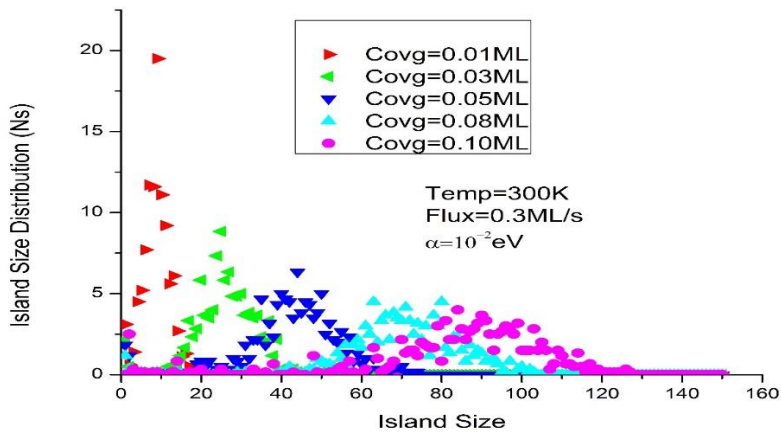

Fig. 4.5.8: Graph of the island size distribution(Ns) vs. island size(S) for flux 0.3 ML/s at T=300K for pattern impurity with strain-driven detachment and inclusion of strain monomer diffusion with strain parameter 0.01eV. The simulation parameters are L=300, flux=0.3ML/s, impurity separation=30a where "a" is atomic distance, number of impurity=100, Es=0.4eV, En=0.1eV, Ed=0.6eV, α =0.01eV.

# CHAPTER 5

## 5.1. SUMMARY AND CONCLUSIONS

In this work, we have studied the effect of a pattern impurity with varying strain-driven detachment and inclusion of strain monomer diffusion during epitaxial growth by using Kinetic Monte Carlo simulation. In this approach, the surface adatoms occupy discrete positions on a lattice and the thermally activated kinetic processes obey Arrhenius dynamics. The activation energy for adatom diffusion consists of four terms: the adatom-substrate interaction, the nearest neighbor interaction between the adatoms, the strain energy due to lattice mismatch and the detachment energy barrier. In order to study the effect of a patterned substrate, the lattice is divided into square-shaped domains. Preferential nucleation site is selected at the center of the domains. The pre-deposited impurity at this site is considered as the nucleation site for island growth. We note that the growth pattern in patterned impurity surface with varying strain-driven detachment and inclusion of strained monomer diffusion at different coverage and temperature.

At higher flux values (i.e. at 0.08ML/s and 0.3ML/s), peak of the distribution curve shifts towards the higher value of S with more height which is due to further growth of impurity nucleating islands for the system existence of strain assisted kinetic mechanism responsible for the detachment events of adatom and strain monomer diffusion is responsible for more size ordering of the growth impurity nucleating island. The growth of impurity nucleating islands with more size ordering yields, greater the value of average island size and lower the value of relative width distribution. Thus, even at high flux values, the effect of patterned impurity on the surface with varying strain-driven detachment contributes for the growth of impurity nucleating islands with more size ordering.

## 5.2. SUGGESTIONS TO FUTURE WORK

The following would be the interesting challenges for future work:

1. To study the effect of patterned impurity on the surface with variation of strian energy for strain-driven detachment of adatoms according to distance from center of mass of islands yields, self-assembling quantum dots (SAQDs).

2. To include the detachment of the dimmer and trimmer from the island.

3. To include the diffusion of adatom single bonded with impurity atom.

4. To model the crystal growth by considering the mobility of impurity atoms.

# References;